\documentclass[useAMS,usenatbib]{mn2e_personal}

\usepackage{graphicx}

\title[EoS and the Riemann problem for ideal flows]{An approach to the Riemann problem in the light of a reformulation of the state equation for SPH inviscid ideal flows: a highlight on spiral hydrodynamics in accretion discs}
\author[G. Lanzafame]{G. Lanzafame\thanks{E-mail:
glanzafame@oact.inaf.it}\\
INAF - Osservatorio Astrofisico di Catania, Via S. Sofia
              78 - 95123 Catania, Italy\\}
\begin{document}

\date{Accepted -------. Received -------; in original form -------}

\pagerange{\pageref{firstpage}--\pageref{lastpage}} \pubyear{2009}

\maketitle

\label{firstpage}

\begin{abstract}
  In physically inviscid fluid dynamics, "shock capturing" methods adopt either an artificial viscosity contribution or an appropriate Riemann solver algorithm. These techniques are necessary to solve the strictly hyperbolic Euler equations if flow discontinuities (the Riemann problem) are to be solved. A necessary dissipation is normally used in such cases. An explicit artificial viscosity contribution is normally adopted to smooth out spurious heating and to treat transport phenomena. Such a treatment of inviscid flows is also widely adopted in the Smooth Particle Hydrodynamics (SPH) finite volume free Lagrangian scheme. In other cases, the intrinsic dissipation of Godunov-type methods is implicitly useful. Instead "shock tracking" methods normally use the Rankine-Hugoniot jump conditions to solve such problems. A simple, effective solution of the Riemann problem in inviscid ideal gases is here proposed, based on an empirical reformulation of the equation of state (EoS) in the Euler equations in fluid dynamics, whose limit for a motionless gas coincides with the classical EoS of ideal gases. The application of such an effective solution to the Riemann problem excludes any dependence, in the transport phenomena, on particle smoothing resolution length $h$ in non viscous SPH flows. Results on 1D shock tube tests, as well as examples of application for 2D shear flows are here shown. As an astrophysical application, a much better identification of spiral structures in accretion discs in a close binary (CB), as a result of this reformulation is also shown here.
\end{abstract}


%
%
%

\section{Introduction}

  In both Lagrangian and Eulerian inviscid fluid dynamics, a dissipation is normally adopted to handle discontinuities in the flow (the Riemann problem). An artificial viscosity is introduced in SPH, as a shock capturing method, to prevent particle interpenetration and to smooth out spurious heating in the flow to solve the strictly hyperbolic system of Euler equations. The introduction of such a small dissipation is also currently adopted to produce both mass and angular momentum transport in SPH physically inviscid modelling of accretion discs in astrophysics \citep{a1,a6,a2,a3,a4,a12,a13,a5,a14,a10,a9,a7,a8,a11}. Efforts were accomplished in SPH to solve both the "approximate" and the "exact" Riemann problem, either via an explicit reformulation of the artificial viscosity term \citep{a15,a17,a16} or via sophisticated Godunov algorithms \citep{a18,a46,a21,a20,a45,a22} which also include an "intrinsic" implicit dissipation. In the first case, a reformulation of the artificial viscosity could be necessary because, for "weak shocks" or low Mach numbers, the fluid becomes "too viscous" and angular momentum and vorticity could be non-physically transferred. An alternative physical way to solve the Riemann problem, based on a reformulation of the EoS in the Euler equations, is here presented, where particle SPH pressure terms are recalculated without any artificial viscosity contribution.

  In this paper we do not solve the Riemann problem by proposing a numerical algorithm, i.e. a Riemann solver. Instead we try to solve the Riemann problem, invalidating the stability of solutions of the hyperbolic Euler equations of inviscid flows, in a strictly physical sense, searching for an EoS for ideal gases including the correct dissipative contribution. This means that, even though we work in SPH formulation, making the most of our experience, final results involve the EoS in general. Since shock flows are non equilibrium events, we pay attention that the EoS: $p = (\gamma - 1) \rho \epsilon$ for ideal flows cannot exactly be applied to solve the Riemann problem. In fact such an equation is strictly valid only for equilibrium or for quasi-equilibrium thermodynamic states, whenever the velocity of propagation of perturbations equals the sound velocity. Successful results, based on some form of mathematical dissipation introduced within the strictly hyperbolic system of non linear Euler equations for ideal flows, are obtained due to the necessity to correct such a congenital defect. In \S 2 of this paper, we briefly recall how the SPH method works for ideal non viscous flows. In the same \S 2 we also outline which evolution has been accomplished in the explicit artificial viscosity dissipation description. Instead, in \S 3, we show how to reformulate the EoS, according to the Riemann problem for inviscid, ideal gases. Analytical formulations of how a physical dissipation in non viscous flows for irreversible physical processes plays on thermodynamic properties is shown in \S 4, according to statistical thermodynamics. In \S 5 a refinement of the same EoS is also formulated to handle a dissipation that is also right for shear flows. Applications to 1D shock tubes \citep{a32} and to blast waves, to solve shocks, to 2D turbulence and 2D shear flows, are also shown in \S 6, as well as a comparison to models adopting different dissipation that is also given for the coming out of spiral patterns in accretion discs in a CB (see App. A).

%
%
%

\section{The Euler equations and their SPH formulation}

\subsection{SPH and artificial viscosity for non viscous flows}

  In the Lagrangian ideal non viscous gas hydrodynamics, the relevant equations (Euler equations) are:

\begin{equation}
\frac{d\rho}{dt} + \rho \nabla \cdot \bmath{v} = 0 \hfill \mbox{continuity equation}
\end{equation}

\begin{equation}
\frac{d \bmath{v}}{dt} = - \frac{\nabla p}{\rho} + \bmath{\Phi} \ \ \ \ \ \ \ \hfill \mbox{momentum equation}
\end{equation}

\begin{equation}
\frac{d \epsilon}{dt} = - \frac{p}{\rho} \nabla \cdot \bmath{v} \hfill \mbox{energy equation}
\end{equation}

\begin{equation}
p = (\gamma - 1) \rho \epsilon \hfill \mbox{perfect gas equation}
\end{equation}

\begin{equation}
\frac{d \bmath{r}}{dt} = \bmath{v} \hfill \mbox{kinematic equation}
\end{equation}

  The majority of the adopted symbols have the usual meaning: $d/dt$ stands for the Lagrangian derivative, $\rho$ is the gas density, $\epsilon$ is the thermal energy per unit mass, $\bmath{\Phi}$ is the external effective force field, also taking into account of centrifugal and Coriolis contributions. The adiabatic index $\gamma$ has the meaning of a numerical parameter whose value lies in the range between $1$ and $5/3$, in principle.

 The SPH method is a Lagrangian scheme that discretizes the fluid into moving interacting and interpolating domains called "particles". All particles move according to pressure and body forces. The method makes use of a Kernel $W$ useful to interpolate a physical quantity $A(\bmath{r})$ related to a gas particle at position $\bmath{r}$ according to \citep{a23,a24}:

\begin{equation}
A(\bmath{r}) = \int_{D} A(\bmath{r}') W(\bmath{r}, \bmath{r}', h) d \bmath{r}'
\end{equation}

$W(\bmath{r}, \bmath{r}', h)$, the interpolation Kernel, is a continuous function - or two connecting continuous functions whose derivatives are continuous even at the connecting point - defined in the spatial range $2h$, whose limit for $h \rightarrow 0$ is the Dirac delta distribution function. All physical quantities are described as extensive properties smoothly distributed in space and computed by interpolation at $\bmath{r}$. In SPH terms we write:

\begin{equation}
A_{i} = \sum_{j=1}^{N} \frac{A_{j}}{n_{j}} W(\bmath{r}_{i}, \bmath{r}_{j}, h) = \sum_{j=1}^{N} \frac{A_{j}}{n_{j}} W_{ij}
\end{equation}

where the sum is extended to all particles included within the interpolation domain $D$, $n_{j} = \rho_{j}/m_{j}$ is the number density relative to the $jth$ particle. $W(\bmath{r}_{i}, \bmath{r}_{j}, h) \leq 1$ is the adopted interpolation Kernel whose value is determined by the relative distance between particles $i$ and $j$. Typically, such cubic spline Kernels $W(|\bmath{r_{ij}}|,h)$ are in the form:

\begin{equation}
W_{ij} = \frac{1}{\pi h^{3}} \left\{ \begin{array}{ll}
1 - \frac{3}{2} |\bmath{r_{ij}}|^{2} + \frac{3}{4} |\bmath{r_{ij}}|^{3} & \textrm{if $0 \leq |\bmath{r_{ij}}| \leq 1$}\\
\frac{1}{4} (2 - |\bmath{r_{ij}}|)^{3} & \textrm{if $1 \leq |\bmath{r_{ij}}| \leq 2$}\\
0 & \textrm{otherwise,}
\end{array} \right.
\end{equation}

even though also Gaussian forms are adopted. In expression (8), $|\bmath{r_{ij}}| = |\bmath{r}_{i} - \bmath{r}_{j}|/h$ represents the module of the radial distance between particles $i$ and $j$ in units of $h$. Other SPH Kernel analytical formulations could be considered \citep{a47}. However this is beyond the aim of this paper because the physical solution of the Riemann problem to solve the strictly hyperbolic Euler equations of fluid dynamics does not rely on the Kernel choice related to the numerical scheme. The Kernel formulation (8) is anyway the most popular since the 90's \citep{a24}.

  In SPH formalism, equations (2) and (3) take the form, respectively:
  
\begin{equation}
\frac{d \bmath{v}_{i}}{dt} = - \sum_{j=1}^{N} m_{j} \left( \frac{p_{i}}{\rho_{i}^{2}} + \frac{p_{j}}{\rho_{j}^{2}} \right) \nabla_{i} W_{ij} + \bmath{\Phi}_{i}
\end{equation}

\begin{equation}
\frac{d \epsilon_{i}}{dt} = \frac{1}{2} \sum_{j=1}^{N} m_{j} \left( \frac{p_{i} }{\rho_{i}^{2}} + \frac{p_{j}}{\rho_{j}^{2}}\right)  \bmath{v}_{ij} \cdot \nabla_{i} W_{ij}
\end{equation}

where $\bmath{v}_{ij} = \bmath{v}_{i} - \bmath{v}_{j}$ and $m_{j}$ is the mass of $jth$ particle.

  For a better energy conservation, the total energy $E = (\epsilon + \frac{1}{2} v^{2})$ can also be introduced in the SPH formulation:

\begin{equation}
\frac{d}{dt} E_{i} = - \sum_{j=1}^{N} m_{j} \left( \frac{p_{i} \bmath{v}_{i}}{\rho_{i}^{2}} + \frac{p_{j} \bmath{v}_{j}}{\rho_{j}^{2}} \right) \cdot \nabla_{i} W_{ij} + \bmath{\Phi}_{i} \cdot \bmath{v}_{i}
\end{equation}

of the energy equation:

\begin{equation}
\frac{d}{dt} \left( \epsilon + \frac{1}{2} v^{2}\right) = - \frac{1}{\rho} \nabla \cdot \left( p \bmath{v} \right) + \bmath{\Phi} \cdot \bmath{v}.
\end{equation}

In this scheme the continuity equation takes the form:

\begin{equation}
\frac{d\rho_{i}}{dt} = \sum_{j=1}^{N} m_{j} \bmath{v}_{ij} \cdot \nabla_{i} W_{ij}
\end{equation}

or, as we adopt, it can be written as:

\begin{equation}
\rho_{i} = \sum_{j=1}^{N} m_{j} W_{ij}
\end{equation}

which identifies the natural space interpolation of particle densities according to equation (7).

  The pressure term also includes the artificial viscosity contribution given by \citet{a23,a24} and \citet{a25}, with an appropriate thermal diffusion term which reduces shock fluctuations. It is given by:

\begin{equation}
\eta_{ij} = \alpha \mu_{ij} + \beta \mu_{ij}^{2},
\end{equation}

where

\begin{equation}
\mu_{ij} = \left\{ \begin{array}{ll}
\frac{2 h \bmath{v}_{ij} \cdot \bmath{r}_{ij}}{(c_{si} + c_{sj}) (|\bmath{r_{ij}}|^{2} + \xi^{2})} & \textrm{if $\bmath{v}_{ij} \cdot \bmath{r}_{ij} < 0$}\\
\\
0 & \textrm{otherwise}
\end{array} \right.
\end{equation}

with $c_{si}$ being the sound speed of the $ith$ particle, $\bmath{r}_{ij} = \bmath{r}_{i} - \bmath{r}_{j}$, $\xi^{2} \ll h^{2}$, $\alpha \approx 1$ and $\beta \approx 2$. These $\alpha$ and $\beta$ parameters of the order of the unity \citep{a26} are usually adopted to damp oscillations past high Mach number shock fronts developed by non linear instabilities \citep{a27}. SPH method, like other finite difference schemes, is far from the continuum limit. The linear $\alpha$ term is based on the viscosity of a gas. The quadratic ($\beta$, Von Neumann-Richtmyer-like) artificial viscosity term is necessary to handle strong shocks. In the physically inviscid SPH gas dynamics, angular momentum transport is mainly due to the artificial viscosity included in the pressure terms as:

\begin{equation}
\frac{p_{i}}{\rho_{i}^{2}} + \frac{p_{j}}{\rho_{j}^{2}} = \left( \frac{p_{i}}{\rho_{i}^{2}} + \frac{p_{j}}{\rho_{j}^{2}} \right)_{gas} (1 + \eta_{ij})
\end{equation}

where terms into parentheses refer to intrinsic gas properties.

  In SPH conversion (eqs. 9, 10, 11, 13, 14) of mathematical equations (eqs. 1 to eq. 3) there are two principles embedded. Each SPH particle is an extended, spherically symmetric domain where any physical quantity $f$ has a density profile $f W(\bmath{r}_{i}, \bmath{r}_{j}, h) \equiv f W(|\bmath{r}_{i} - \bmath{r}_{j}|, h)$. Besides, the fluid quantity $f$ at the position of each SPH particle could be interpreted by filtering the particle data for $f(\bmath{r})$ with a single windowing function whose width is $h$. So doing, fluid data are considered isotropically smoothed all around each particle along a length scale $h$. Therefore, according to such two concepts, the SPH value of the physical quantity $f$ is both the superposition of the overlapping extended profiles of all particles and the overlapping of the closest smooth density profiles of $f$.

%
%
%

\subsection{The Riemann problem in the SPH formulation}

  Strong shock require $\alpha = 1$. However for weak shocks and for small Mach number flows, the fluid becomes "too viscous" and both angular momentum and vorticity are transferred unphysically. To solve this problem, several solutions are proposed:

  a) the formulation of a "limiter" \citep{a15}, multiplying the artificial viscosity terms $\eta_{i}$ for $f_{ij} = 0.5 (f_{i} + f_{j})$, where

\begin{equation}
f_{i} = \frac{|\nabla \cdot \bmath{v_{i}}|}{|\nabla \cdot \bmath{v_{i}}| + |\nabla \times \bmath{v_{i}}| + 10^{-4} c_{si}/h}.
\end{equation}

  It reduces the unphysical spread of angular momentum in whirling flows up to $20$ times. It is $\approx 1$ for planar shocks, while it increases if the tangential kinematics is relevant;

  b) a switch to reduce artificial viscosity \citep{a16}. In this hypothesis, for each $ith$ particle the $\alpha$ evolves according to a decay equation:

\begin{equation}
\frac{d \alpha_{i}}{dt} = - \frac{\alpha_{i} - \alpha^{\ast}}{\tau_{i}} + S_{i},
\end{equation}

where $\alpha^{\ast} \approx 0.1$, $\tau_{i} = h/c_{s}$, $c_{s}$ is the sound speed, and the source term $S_{i} = f_{i} \max(- \nabla \cdot \bmath{v}_{i}, 0)$.

  c) a reformulation of the artificial viscosity according to the Riemann problem \citep{a17}. Particles $i$ and $j$ are considered as the equivalent left "$l$" and right "$r$" states of a given contact interface. The 1D Riemann problem is taken into account along the line joining them. Being the Euler equations in conservative form:

\begin{equation}
\frac{\partial \bmath{s}}{\partial t} + \frac{\partial \bmath{f}}{\partial x} = 0,
\end{equation}

the simplest Euler technique of integration is:

\begin{equation}
\bmath{s}_{i}^{n+1} = \bmath{s}_{i}^{n} - \frac{\Delta t}{\Delta x} \left[ \bmath{f}^{\ast}(\bmath{s}_{i}, \bmath{s}_{i-1}) - \bmath{f}^{\ast}(\bmath{s}_{i-1}, \bmath{s}_{i}) \right],
\end{equation}

where numerical fluxes \citep{a31}

\begin{equation}
\bmath{f}^{\ast}(\bmath{s}_{l}, \bmath{s}_{r}) = 0.5 \left( \bmath{f}_{l}^{\ast} + \bmath{f}_{r}^{\ast} - \sum_{j=1}^{3} \Big| \lambda_{j}^{\ast} \Delta \omega_{j}^{\ast} \bmath{e}_{j} \Big| \right),
\end{equation}

where the $\bmath{e}_{j}$ are the eigenvectors of the Jacobian matrix $\bmath{A} = \partial \bmath{f}/\partial \bmath{s}$ and $\lambda_{j}^{\ast}$ is an average of $\lambda$ for the "$l$" and "$r$" states. $\Delta \omega_{j}^{\ast}$ are the "jumps" of $\bmath{s}$ across the characteristics:

\begin{equation}
\bmath{s}_{r} - \bmath{s}_{l} = \sum_{j=1}^{3} \Delta \omega_{j}^{\ast} \bmath{e}_{j}.
\end{equation}

  For 1D ideal flows, the eigenvalues are $v$, $v + c_{s}$ and $v - c_{s}$, where the two including the sound velocity are the velocities of propagation of sound referred to the frame whose fluid velocity is $v$. Assuming that the jump in the velocity across characteristics could physically be $\bmath{v}_{ij} \cdot \bmath{r}_{ij}/|\bmath{r}_{ij}|$ and that a signal velocity $v_{sig}$ corresponds to the above eigenvalues, $|\lambda_{j}^{\ast}| \Delta \omega_{j}^{\ast}$ corresponds to $v_{sig, ij} \bmath{v}_{ij} \cdot \bmath{r}_{ij}/|\bmath{r}_{ij}|$. So doing, the artificial pressure contribution in the momentum equation is:

\begin{equation}
\left( \frac{p_{i}}{\rho_{i}^{2}} + \frac{p_{j}}{\rho_{j}^{2}} \right)_{gas}  \eta_{ij} = - \frac{K v_{sig, ij} \bmath{v}_{ij} \cdot \bmath{r}_{ij}/|\bmath{r}_{ij}|}{\rho_{ij}},
\end{equation}

where $K$ is an arbitrary constant $\approx 1$ and $\rho_{ij} = 0.5 (\rho_{i} + \rho_{j})$.

  As far as the energy equation is concerned, the SPH formulation in this case is:

\begin{eqnarray}
\frac{d}{dt} E_{i} & = & - \sum_{j=1}^{N} m_{j} \left( \frac{p_{i} \bmath{v}_{i}}{\rho_{i}^{2}} + \frac{p_{j} \bmath{v}_{j}}{\rho_{j}^{2}} \right)_{gas} \cdot \nabla_{i} W_{ij}  \nonumber \\
& & - \frac{K v_{sig, ij} e_{ij}^{\ast} \bmath{r}_{ij}/|\bmath{r}_{ij}|}{\rho_{ij}} \cdot \nabla_{i} W_{ij} + \bmath{\Phi}_{i} \cdot \bmath{v}_{i},
\end{eqnarray}

being $e_{ij}^{\ast} = e_{i}^{\ast} - e_{j}^{\ast}$, where $e_{i}^{\ast} = 0.5 (\bmath{v}_{i} \cdot \bmath{r}_{ij}/|\bmath{r}_{ij}|)^{2} + \epsilon_{i}$.

  The signal velocities for the 1D Lagrangian Riemann problem on ideal flows are reported in \citep{a30}, on the basis of \citet{a28} and \citet{a29} results. The pressure on the contact interface $p^{\ast}$ links the left and the right states. For systems with one shock and one rarefaction wave $p^{\ast} \in [p_{l}, p_{r}]$ and

\begin{equation}
p^{\ast} = \left[ \frac{c_{sl} + c_{sr} + (v_{l} - v_{r}) (\gamma - 1)/2}{c_{sl}/p_{l}^{(\gamma - 1)/2 \gamma} + c_{sr}/p_{r}^{(\gamma - 1)/2 \gamma}} \right]^{2 \gamma/(\gamma - 1)}.
\end{equation}

In the case of two shocks, a more complicated relation \citep{a28} also involves the velocity $v^{\ast}$ on the contact interface. However, for some practical purposes, $p^{\ast} = \max(p_{l}, p_{r})$ and $v^{\ast} = 0.5 (v_{l} + v_{r})$. Notice that in the presence of two very strong rarefaction waves $p^{\ast} = 0$.

  The two Lagrangian velocities of waves spawned by the interface are:

\begin{equation}
\lambda_{l} = \left\{ \begin{array}{ll}
v_{l} - c_{sl} [1 + \frac{(\gamma - 1) (p^{\ast}/p_{l} - 1)}{2 \gamma}] & \textrm{if $p^{\ast}/p_{l} > 1$}\\
\\
v_{l} - c_{sl} & \textrm{if $p^{\ast}/p_{l} \leq 1$}
\end{array} \right.
\end{equation}

and

\begin{equation}
\lambda_{r} = \left\{ \begin{array}{ll}
v_{r} + c_{sr} [1 + \frac{(\gamma - 1) (p^{\ast}/p_{r} - 1)}{2 \gamma}] & \textrm{if $p^{\ast}/p_{r} > 1$}\\
\\
v_{r} + c_{sr} & \textrm{if $p^{\ast}/p_{r} \leq 1$}
\end{array} \right.
\end{equation}

In the Lagrangian description, being $v_{sig}$ the speed of propagation of perturbation from $i$ to $j$ and vice-versa ($l \leftrightarrow r$),

\begin{eqnarray}
v_{sig, i \rightarrow j} & = & c_{si} - \bmath{v}_{i} \cdot \bmath{r}_{ij}/|\bmath{r}_{ij}| \\
v_{sig, j \rightarrow i} & = & - c_{sj} - \bmath{v}_{j} \cdot \bmath{r}_{ij}/|\bmath{r}_{ij}| \\
v_{sig} & = & v_{sig, i \rightarrow j} - v_{sig, j \rightarrow i} \\
& = & c_{si} + c_{sj} - \bmath{v}_{ij} \cdot \bmath{r}_{ij}/|\bmath{r}_{ij}|
\end{eqnarray}

having considered the versus going from $i$ to $j$. By performing some numerical experiments, \citet{a17} also suggested other similar algebraic expressions.

\subsection{Some cautions in using artificial viscosities}

  In \citet{a33}, some cautionary remarks are reported on the adoption of artificial dissipation. In particular, "Because the magnitude of the viscous term does not affect the net shock jump conditions, many numerical schemes implicitly or explicitly incorporate the trick of {\it artificial viscosity} for halting the ever-growing steepening tendency produced by nonlinear effects, thereby gaining the automatic insertion of shock waves wherever they are needed (in time-dependent calculations)." However, the numerical viscous term should be large enough to spread out shock transitions only over a few resolution lengths, making shock waves resolvable. In this sense, any mathematical dissipation should be considered a useful mathematical "trick" to get correct results.

%
%
%

\section{How EoS matches the Riemann problem}

  The solution of the Riemann problem is obtained at the interparticle contact points among particles, where a pressure and a velocity, relative to the flow discontinuity, are computed. This is also clearly shown in \citet{a46,a21,a20,a45,a22}, where the new pressure $p^{\ast}$ and velocity $v^{\ast}$ are reintroduced in the Euler equations instead of $p$ and $v$ to obtain the new solutions compatible with inviscid flow discontinuities. In SPH, we pay attention in particular \citep{a46,a21,a20,a45} to the particle pressures $p_{i}$ and $p_{j}$, in the SPH formulation of the momentum and energy equations (9) and (10 or 11), whose substitution with pressures $p_{i}^{\ast}$ and $p_{j}^{\ast}$, solutions of the Riemann problem, excludes any artificial viscosity adoption, although a dissipation, implicitly introduced, is necessary. Therefore, the solution of the Lagrangian Riemann problem, only in its pressure computation, is enough to interface SPH with the Riemann problem solution.

  The key concept is that the physical interpretation of the application of the artificial viscosity contribution in the pressure terms corresponds to a reformulation of the EoS for inviscid ideal gases, whose equation:

\begin{equation}
p|_{gas} = (\gamma - 1) \rho \epsilon
\end{equation}

is strictly applied in fluid dynamics when the gas components do not collide with each other. In the case of gas collisions, it modifies in:

\begin{equation}
p^{\star} = (\gamma - 1) \rho \epsilon + \textrm{other.}
\end{equation}

  The further term takes into account the velocity of perturbation propagation \citep{a17}. This velocity equals the ideal gas sound velocity $c_{s}$ whenever we treat non collisional gases in equilibrium or in the case of rarefaction waves. Instead, it includes the "compression velocity": $\bmath{v}_{ij} \cdot \bmath{r}_{ij}/|\bmath{r}_{ij}|$ in the case of shocks (eq. 29). In the first case, we write the EoS for inviscid ideal gases as:

\begin{equation}
p|_{gas} = \frac{\rho}{\gamma} c_{s}^{2},
\end{equation}

where $c_{s} = (\gamma p/\rho)^{1/2} = [\gamma (\gamma - 1) \epsilon]^{1/2}$. Instead, in the second case, the new formulation for the EoS is obtained squaring $v_{sig, i \rightarrow j}$, so that $c_{s}^{2} (1 - v_{shock}/c_{s})^{2}$ is an energy per unit mass in the case of shock. Hence:

\begin{equation}
p^{\star} = \left\{ \begin{array}{ll}
\frac{\rho}{\gamma} c_{s}^{2} \left(1 - \frac{v_{shock}}{c_{s}} \right)^{2} & \textrm{if $v_{shock} < 0$}\\
\\
\frac{\rho}{\gamma} c_{s}^{2} & \textrm{if $v_{shock} \ge 0$}
\end{array} \right.
\end{equation}

  In the SPH scheme, being:

\begin{equation}
p_{i}^{\star} = \frac{\rho_{i}}{\gamma} c_{si}^{2} \left(1 - \frac{v_{shock,i}}{c_{si}} \right)^{2},
\end{equation}

\begin{equation}
v_{shock,i} = \left\{ \begin{array}{ll}
\frac{\bmath{v}_{ij} \cdot \bmath{r}_{ij}}{|\bmath{r}_{ij}|} & \textrm{if $\bmath{v}_{ij} \cdot \bmath{r}_{ij} < 0$}\\
\\
0 & \textrm{otherwise.}
\end{array} \right.
\end{equation}

  This formulation introduces the "shock pressure term": $\rho (v_{shock}^{2} - 2 v_{shock} c_{s})/\gamma$, whose linear and quadratic power dependence on $\bmath{v}_{ij} \cdot \bmath{r}_{ij}/|\bmath{r}_{ij}|$ is analogue to both the linear and the quadratic components of the artificial viscosity terms (15). The linear term $\propto c_{s} v_{shock}$ is based on the viscosity of a gas. The quadratic term $\propto (c_{s} v_{shock})^{2}$ (Von Neumann-Richtmyer-like) is necessary to handle strong shocks. These contributions involve a dissipative power, whose effect corresponds to an increase of the gas pressure. Therefore, we adopt the formulation (eq. 21) as $p_{i}^{\star}$ and $p_{j}^{\star}$ in the SPH formulation of the momentum (eq. 9) and energy equations (eqs. 10 or 11):

\begin{equation}
\frac{d \bmath{v}_{i}}{dt} = - \sum_{j=1}^{N} m_{j} 
\left( \frac{p_{i}^{\star}}{\rho_{i}^{2}} + \frac{p_{j}^{\star}}{\rho_{j}^{2}} \right) \nabla_{i} W_{ij} + \bmath{\Phi}_{i}
\end{equation}

\begin{equation}
\frac{d \epsilon_{i}}{dt} = \frac{1}{2} \sum_{j=1}^{N} m_{j} \left( \frac{p_{i}^{\star} }{\rho_{i}^{2}} + \frac{p_{j}^{\star}}{\rho_{j}^{2}}\right) \bmath{v}_{ij} \cdot \nabla_{i} W_{ij},
\end{equation}

\begin{equation}
\frac{d}{dt} E_{i} = - \sum_{j=1}^{N} m_{j} \left( \frac{p_{i}^{\star} \bmath{v}_{i}}{\rho_{i}^{2}} + \frac{p_{j}^{\star} \bmath{v}_{j}}{\rho_{j}^{2}} \right) \cdot \nabla_{i} W_{ij} + \bmath{\Phi}_{i} \cdot \bmath{v}_{i}.
\end{equation}

  This simple reformulation, allows us to keep the same Courant-Friedrichs-Lewy condition as for the timestep computation, substituting only the sound velocity $c_{s}$ with $c_{s} - v_{shock}$.

  Notice that the comparison of physically dissipative shock pressure terms $\rho (v_{shock}^{2} - 2 v_{shock} c_{s})/\gamma$ with those relative to SPH artificial viscosity: $p_{i}|_{gas} \eta_{ij}$ gives for $\alpha$ and $\beta$ parameters the physical equivalence for the $ith$ particle:

\begin{eqnarray}
\alpha & = & 2 \frac{c_{sij} |\bmath{r}_{ij}|}{c_{si} h} \\
\beta & = & \frac{c_{sij}^{2} |\bmath{r_{ij}}|^{2}}{c_{si}^{2} h^{2}} \ = \ \left( \frac{\alpha}{2} \right)^{2},
\end{eqnarray}

where $c_{sij} = 0.5 (c_{si} + c_{sj})$.

  The linear and the quadratic dependence of $\alpha$ and $\beta$ on $c_{sij} |\bmath{r}_{ij}|/(c_{si} h)$ perfectly correlates the physical dissipation of a perfect gas both to a bulk viscosity and to a quadratic Von Neumann-Richtmyer-like dissipation, able to handle strong shocks, avoiding any dependence on $h$, contrary to SPH artificial viscosity. In fact, both $h$ and $h^{ij}$ algebrically erase applying eqs. (15) and (16). Compared to SPH constant $\alpha \sim 1$ and $\beta \sim 2$, such correlations show that $\alpha$ and $\beta$ equal zero for $|\bmath{r_{ij}}| = 0$ and that $\alpha = 2$ and $\beta = 1$ in a homogeneous and isotropic gas for particle mutual separations $|\bmath{r_{ij}}| = h$. Both parameters linearly and quadratically variate, respectively, varying either $|\bmath{r}_{ij}|/h$, or $c_{sij}/c_{si}$, or both. Hence, both $\alpha$ and $\beta$ decrease for closer particle neighbours and/or for colder companions, whose $c_{sj}$ is smaller than $c_{si}$.

%
%
%

\section{Consequences of a shock physical dissipation on the Maxwell-Boltzmann statistical thermodynamics}

  Thermodynamic properties of a system in thermal equilibrium are described by the partition function \citep{a39,a40} $Z$:

\begin{eqnarray}
Z & = & \sum_{i}^{energy \ levels} G_{i} e^{- \beta U_{i}} = \sum_{j}^{quantum \ states} e^{- \beta U_{j}} \nonumber \\
& = & \sum_{j}^{complexions} e^{- \beta U_{j}}.
\end{eqnarray}

  The most probable distribution of systems in the ensemble with energies $U_{i}$ and with quantum states of the system, given by the Maxwell-Boltzmann law, are:

\begin{equation}
\overline{n}_{i} = G_{i} e^{-\alpha} e^{-\beta U_{i}}
\end{equation}

 and

\begin{equation}
\overline{m}_{j} = e^{-\alpha} e^{-\beta U_{j}}
\end{equation}

respectively, where $e^{-\alpha} = Z/N$, being $N = \sum_{i} n_{i}$ the total number of systems in the ensemble, $U = \sum_{i} n_{i} U_{i}$ the total energy of the ensemble and being $G_{i}$ the degeneracy of the energy level $U_{i}$. The parameter $\beta$ must be strictly positive to have $Z$ meaningful. $\beta = (K_{B} T)^{-1}$, where $K_{B}$ is the Boltzmann constant and $T$ is the temperature. Being $U_{i}$ classically proportional to the square of the linear momentum: $(U_{i} \propto v_{i}^{2})$ for non interacting free atoms, without considering their internal energy levels, each exponential in the summations is related to a Gaussian distribution function.

  Since in fluid dynamics equilibrium conditions are a special case, the application of thermodynamic laws is an evident approximation, in particular whenever non reversible events occur like shock waves. We assume the hypothesis that even in conditions of a fluid motion, the Maxwell-Boltzmann law holds, where the net effect of any physical dissipation is to reduce the peak value of the Maxwell-Boltzmann distribution, enlarging its spread, conserving both $N$ and $U$. Hence, any physical dissipation is introduced as a $0 < D \leq 1$ term ($D = 1$ involves no dissipation) multiplying $\beta U_{i}$ or $\beta U_{j}$ in the above exponentials as: $\beta D U_{i}$ or $\beta D U_{j}$, so that:

\begin{eqnarray}
Z & = & \sum_{i}^{energy \ levels} G_{i} e^{- \beta D U_{i}} = \sum_{j}^{quantum \ states} e^{- \beta D U_{j}} \nonumber \\
& = & \sum_{j}^{complexions} e^{- \beta D U_{j}}.
\end{eqnarray}

  This is necessary to determine the desired result, without any alteration of the system's degeneracy, which cannot be considered in so far as we are dealing with an ideal gas. In a microscopic description, physical dissipation irreversibly converts ordered kinetic energy into chaotic kinetic energy. That is the conversion of macroscopic kinetic energy into thermal energy, in so far as the thermodynamic system is always in thermal equilibrium, obeying Markovian statistics. Atoms or molecules do not interact with each other, and their internal quantum energies do not affect the global thermodynamics. In the Maxwell-Boltzmann distribution hypothesis, this implies a transition from one Maxwell-Boltzmann statistical distribution to another one, whose half weight at half height is larger. If such a transition occurs during the shock crossing, the whole thermodynamic system could make an instantaneous transition between two thermodynamic equilibrium states. Hence, the larger the half weight at half height of the statistical distribution, the larger the physical dissipation occurred. Thus, writing $\beta^{\star} = D (K_{B} T^{\star})^{-1}$ or, that is the same: $\beta^{\star} = D \beta = D (K_{B} T)^{-1}$,
  
\begin{eqnarray}
Z & = & \sum_{i}^{energy \ levels} G_{i} e^{- \beta^{\star} U_{i}} = \sum_{j}^{quantum \ states} e^{- \beta^{\star} U_{j}} \nonumber \\
& = & \sum_{j}^{complexions} e^{- \beta^{\star} U_{j}}.
\end{eqnarray}

  The net effect of such a dissipation is the desired enlargement of the width at half height of each Gaussian distribution for $Z$, conserving both $N$ and $U$. Thermodynamic properties are obtained \citep{a39,a40} in the same way as they are currently given using $\beta^{\star}$ instead of $\beta$ as:

\begin{itemize}

\item Internal Energy:

\end{itemize}

\begin{equation}
U^{\star} = - \frac{\partial \ln Z/\partial T|_{V}}{\partial \beta^{\star}/\partial T|_{V}} = \left( \frac{K_{B} T}{D} - \frac{1}{\beta} \frac{K_{B} T}{\partial D/\partial T|_{V}} \right) \frac{\partial \ln Z}{\partial T} \Big|_{V}
\end{equation}

keeping fixed the volume $V$;

\begin{itemize}

\item Entropy:

\end{itemize}

  From the Boltzmann law,

\begin{equation}
S^{\star} = - \frac{K_{B} N}{D} \ln Z + \frac{U^{\star}}{T};
\end{equation}

\begin{itemize}

\item Free Helmholtz Energy:

\end{itemize}

\begin{equation}
F^{\star} = U^{\star} - TS^{\star} = - \frac{N K_{B} T}{D} \ln Z;
\end{equation}

\begin{itemize}

\item Pressure:

\end{itemize}

In the hypothesis that dissipation does not explicitly depend on $V$, 

\begin{equation}
p^{\star} = - \frac{\partial F^{\star}}{\partial V}\Big|_{T} = \frac{N K_{B} T}{D} \frac{\ln Z}{V}\Big|_{T}.
\end{equation}

In the classical case of free, non interacting atoms, $\partial \ln Z/\partial V|_{T} = 1/V$, hence

\begin{equation}
p^{\star} = \frac{N}{\beta^{\star}} \frac{1}{V}.
\end{equation}

Thus, being $\beta^{\star} = D \beta = D (K_{B} T)^{-1}$,

\begin{equation}
p^{\star} = N \frac{K_{B} T}{V} \frac{1}{D}.
\end{equation}

  Assuming $D = 1/(1 - v_{shock,i}/c_{si})^{2} \leq 1$; $v_{shock,i} = \bmath{v}_{i,j} \cdot \bmath{r}_{i,j}/|\bmath{r}_{i,j}|$ if $\bmath{v}_{i,j} \cdot \bmath{r}_{i,j} < 0$, $v_{shock,i} = 0$ otherwise; we write an expression that does not depend on volume $V$ and corresponding to a function that, even depending on $T$, it can be easily handled by considering the derivatives in (54). Thus, the EoS is:

\begin{equation}
p^{\star} = \left\{ \begin{array}{ll}
\frac{N K_{B} T}{V} \left(1 - \frac{v_{shock,i}}{c_{si}}\right)^{2} & \textrm{if $v_{shock} < 0$, i.e. $D < 1$}\\
\\
\frac{N K_{B} T}{V} & \textrm{if $v_{shock} \ge 0$, i.e. $D = 1$}
\end{array} \right.
\end{equation}

physically corresponding to the EoS for ideal flows, even taking into account a shock occurrence, exactly in the form (36, or 37+38) previously written. Enthalpy and Gibbs free energy are given by: $H^{\star} = U^{\star} + p^{\star}V$ and $G = U^{\star} + p^{\star}V - TS^{\star}$, respectively.

\section{Which EoS is also right for shear flows?}

  To make a generalization, we need only one general EoS and not a separation of the EoS according to the kinematic of the flow. To this purpose, we can generalize the EoS: $p^{\star} = \rho c^{2}_{s} (1 - v_{shock}/c_{s})^{2}/\gamma$ as:

\begin{equation}
p^{\star} = \frac{\rho}{\gamma} c_{s}^{2} \left(1 - C \frac{v_{R}}{c_{s}} \right)^{2},
\end{equation}

where $C \rightarrow 1$ for $v_{R} = \bmath{v}_{ij} \cdot \bmath{r}_{ij}/|\bmath{r}_{ij}| < 0$, whilst $C \rightarrow 0$ otherwise. A simple empirical formulation can be,

\begin{equation}
C = \frac{1}{\pi} \textrm{arccot} \left( R \frac{v_{R}}{c_{s}} \right),
\end{equation}

where $R \gg 1$. $R$ is a large number describing how much the flow description corresponds to that of an ideal gas. To this purpose, $R \approx \lambda/d$, being $\lambda \propto \rho^{-1/3}$ the molecular mean free path, and $d$ the mean linear dimension of gas molecules. 

  Although the physical meaning of $v_{R}$ in the field of a free Lagrangian particle technique is clear, it could be controversial in an Eulerian description. In this case, the local physical properties of medium have to be taken into account because it is necessary to correlate the approaching of two particles to a local compression. To convert expression (56) to a more general form, we pay attention to the continuity equation regarding the numerical concentration $n$ as:

\begin{equation}
\frac{dn}{dt} + n \nabla \cdot \bmath{v} = 0
\end{equation}

to determine $v_{R}$ as:

\begin{equation}
v_{R} = \frac{dn^{-1/3}}{dt} = - \frac{1}{3} n^{-4/3} \frac{dn}{dt} = \frac{1}{3} n^{-1/3} \nabla \cdot \bmath{v},
\end{equation}

so that eq. (56) can also be written as:

\begin{equation}
p^{\star} = \frac{\rho}{\gamma} c_{s}^{2} \left(1 - C \frac{n^{-1/3} \nabla \cdot \bmath{v}}{3 c_{s}} \right)^{2},
\end{equation}

where

\begin{equation}
C = \frac{1}{\pi} \textrm{arccot} \left( R \frac{n^{-1/3} \nabla \cdot \bmath{v}}{3 c_{s}} \right).
\end{equation}

  EoS in the form of eq. (36, or 37+38, or 56, or 60) are equivalent and effective strictly to solve the Riemann problem. However, if there is a velocity shear in the inviscid ideal flow, the physical dissipation within the EoS (eq. 36, 37+38, or 56) is activated, despite the lack of any shock, whenever close portions of fluid approach each other, showing different densities. This is obviously due to the fact that the physical dissipation in the EoS is activated in the approaching of only two local particles with each other. Looking at the correlation of physical dissipation with $\alpha$ and $\beta$ SPH parameters (eqs. 42, 43), a dissipation still occurs even though dissipation depends on local thermodynamic conditions. This is an initial advantage compared to classical SPH. Such a difficulty, more limited compared to that arising in SPH (or whatever is the technique adopting an explicit artificial dissipation), is totally absent in the form of eq. (60), because it is free of any dissipation whenever $\nabla \cdot \bmath{v} \geq 0$. As an example, any simulation relative to non viscous Keplerian flows, where $\nabla \cdot \bmath{v} = 0$ at the beginning, does not produce any dissipation and fluid motion is stationary endlessly. This is a necessary introduction to the tests shown in the subsequent sections. In its form (60) the reformulated EoS for non viscous ideal gases includes a physical dissipation, depending on general local conditions, which is activated only when a local compression occurs. Hence, dissipation in SPH-based techniques is no longer due to the approaching of only two particles, but to the general physical properties of the local medium, where a lot of particles have to be considered.

\begin{figure*}
\resizebox{\hsize}{!}{\includegraphics[clip=true]{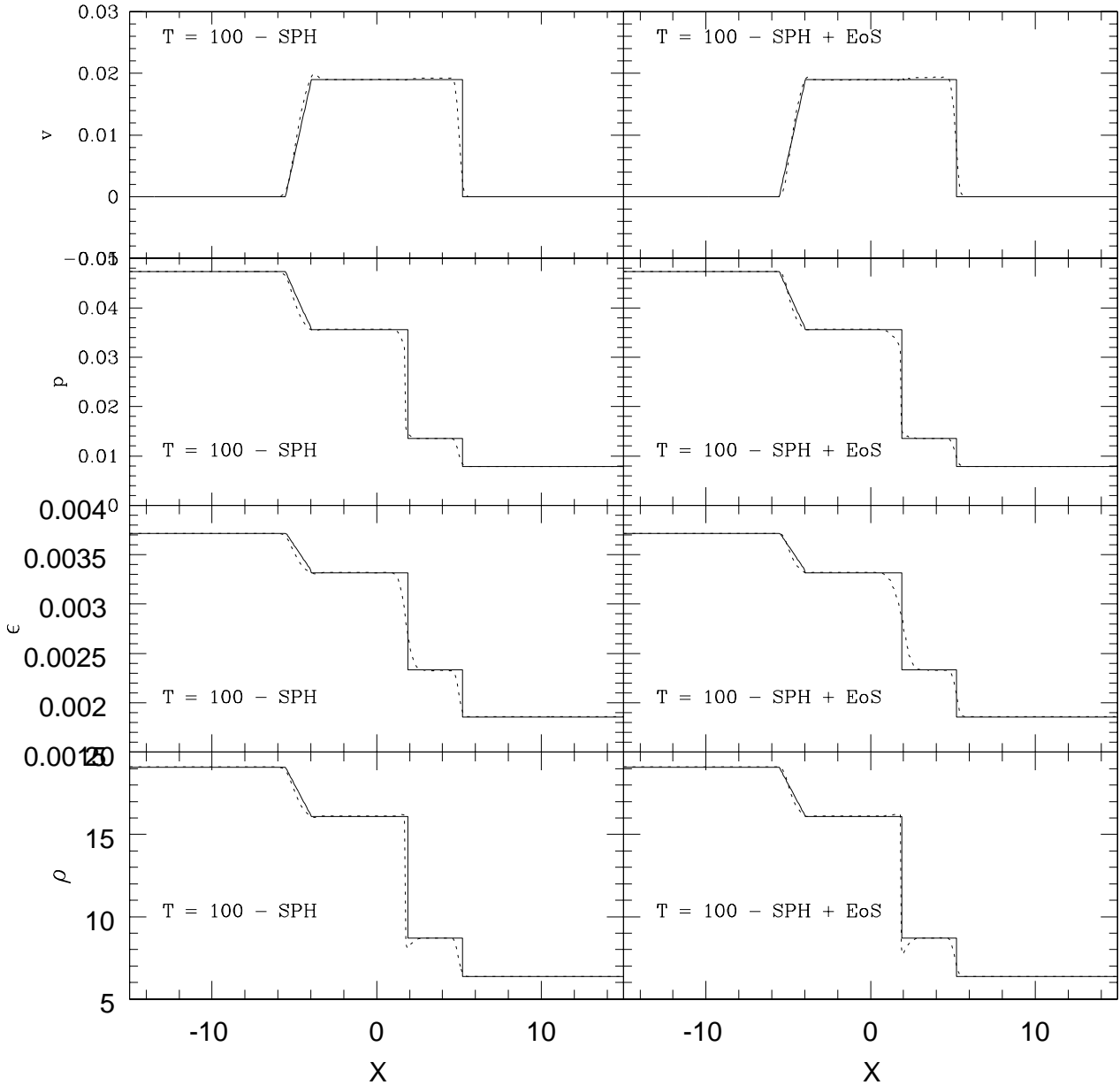}}
\caption{1D shock tube tests as far as both analytical (solid line) and our SPH-Riemann (SPH+EoS, dots) results are concerned (right side plots). Density $\rho$, thermal energy $\epsilon$, pressure $p$ and velocity $v$ are plotted at time $T = 100$. Density and thermal energy of particles initially at rest at time $T = 0$ refer to values plotted at the two edges of each plot. The initial velocity is zero throughout. SPH results are also reported (left side plots).}
\end{figure*}

\begin{figure*}
\resizebox{\hsize}{!}{\includegraphics[clip=true]{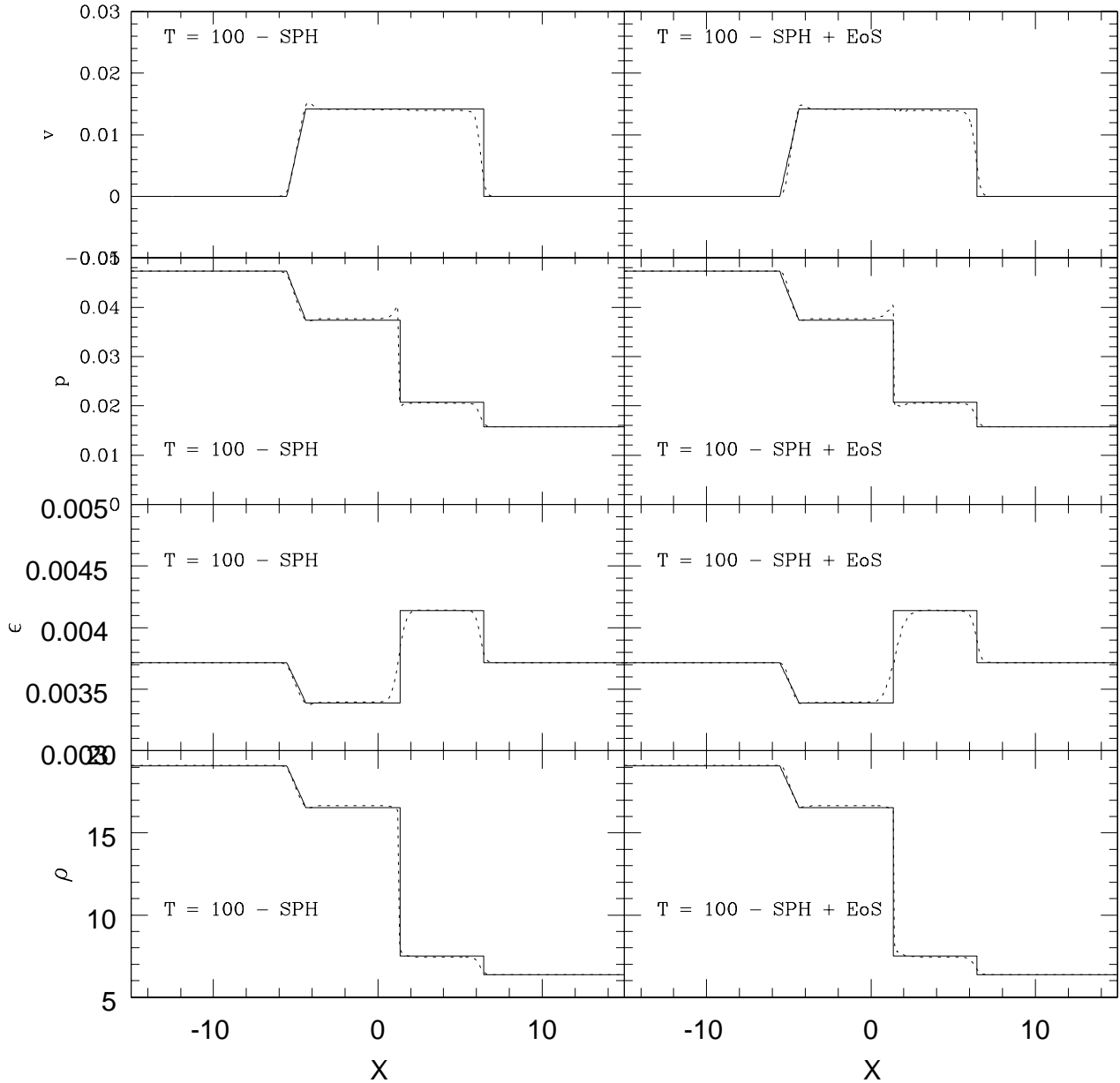}}
\caption{The same as in Fig. 1. In this example, the initial discontinuity does not affect the thermal energy per unit mass, being the gas initially isothermal, while the initial velocity is zero throughout.}
\end{figure*}

  According to such second reformulation of the EoS, also right for shear flows, the algebraic relations determining the $\alpha$ and $\beta$ parameters (similarly as for eqs. 42 and 43) are:

\begin{eqnarray}
\alpha & = & 2 \frac{c_{sij}}{c_{si}} \frac{n_{i}^{- 1/3}}{h} \frac{\nabla \cdot \bmath{v}_{i}}{\bmath{v}_{ij} \cdot \bmath{r}_{ij}} |\bmath{r}_{ij}|^{2} \\
\beta & = &  \frac{c_{sij}^{2}}{c_{si}^{2}} \frac{n_{i}^{- 2/3}}{h^{2}} \left( \frac{\nabla \cdot \bmath{v}_{i}}{\bmath{v}_{ij} \cdot \bmath{r}_{ij}} \right)^{2} |\bmath{r}_{ij}|^{4} \ = \ \left( \frac{\alpha}{2} \right)^{2}.
\end{eqnarray}

  Such relations compare with eqs. (42, 43), respectively, in the case of a pure Riemann problem in fluid dynamics \citep{a75}, being $\nabla \cdot \bmath{v}_{i} \sim \bmath{v}_{ij} \cdot \bmath{r}_{ij} h /|\bmath{r}_{ij}|^{2}$.

\section{Tests}

  In this section we compare solutions in an SPH approach, where the modified EoS is taken into account without any thermal conduction contribution, with analytical solutions, whenever it is possible, as well as with those in SPH where, typically an explicit thermal conductive term is also included \citep{a24}. Tests regard typical flow cases where either collisions (shocks), or turbulence, or shear flows are involved. Throughout the simulations, the initial particle smoothing resolution length is $h = 5 \cdot 10^{-2}$ whilst the dimension of the whole computational domain $D$ is chosen to detect particle turbulence, if it exists, being $D/h \geq 100$ in the most of discussed cases. The adopted $\gamma$ value is $\gamma = 5/3$, whenever not explicitly specified. Results concerning the SPH+EoS approach in the form of eq. (60) will be clearly discussed whenever substantial differences develop.

\subsection{1D Sod shock tubes}

  The behaviour of shock waves moving in the prevailing flow is analytically described by the Rankine-Hugoniot "jump conditions" \citep{a69,a70,a71,a68}. These are obtained by spatially integrating the 1D hyperbolic Euler equations across the discontinuity between the two flow regimes left-right in their Eulerian formalism:

\begin{eqnarray}
\frac{\partial \rho}{\partial t} & = & - \frac{\partial}{\partial x} (\rho v) \\
\frac{\partial \rho v}{\partial t} & = & - \frac{\partial}{\partial x} (\rho v^{2} + p) \\
\frac{\partial \rho E}{\partial t} & = & - \frac{\partial}{\partial x} [\rho v (E + p/\rho)],
\end{eqnarray}

where $E = v^{2}/2 + \epsilon$, whose conservative analytical form can be synthesized as:

\begin{equation}
\frac{\partial w}{\partial t} = - \frac{\partial}{\partial x} f(w).
\end{equation}

In the limit of zero thickness of the shock discontinuity,

\begin{equation}
s (w_{l} - w_{r}) = f(w_{l}) - f(w_{r}).
\end{equation}

Under these conditions a requirement for a unique single-valued solution is that the solution should satisfy the Lax entropy condition \citep{a69,a70,a71} $f'(w_{l}) < s < f'(w_{r})$, where $f'(w_{l})$ and $f'(w_{r})$ are the characteristic speeds at upstream and downstream conditions, respectively. The integrated form of the Rankine-Hugoniot jump conditions are:

\begin{eqnarray}
s (\rho_{l} - \rho_{r}) & = & \rho_{l} v_{l} - \rho_{r} v_{r} \\
s (\rho_{l} v_{l} - \rho_{r} v_{r}) & = & (\rho_{l} v_{l}^{2}) - (\rho_{r} v_{r}^{2}) \\
s (\rho_{l} E_{l} - \rho_{r} E_{r}) & = & \rho_{l} v_{l} E_{l} - \rho_{r} v_{r} E_{r},
\end{eqnarray}

  It is shown, after some algebraic passages, that the shock speed is:
  
\begin{equation}
s = v_{l} + c_{sl} \left[ 1 + \frac{\gamma + 1}{2 \gamma} \left( \frac{p_{r}}{p_{l}} - 1 \right) \right]^{1/2}
\end{equation}

where $c_{sl} = (\gamma p_{l}/ \rho_{l})^{1/2}$. In the case of stationary shocks being both the upstream and downstream pressures positive, there is an upper limit on the density ratio: $\rho_{l}/\rho_{r} \leq (\gamma + 1)/(\gamma - 1)$. However, this limit is currently applied also to non steady shocks.

\begin{figure}
\resizebox{\hsize}{!}{\includegraphics[clip=true]{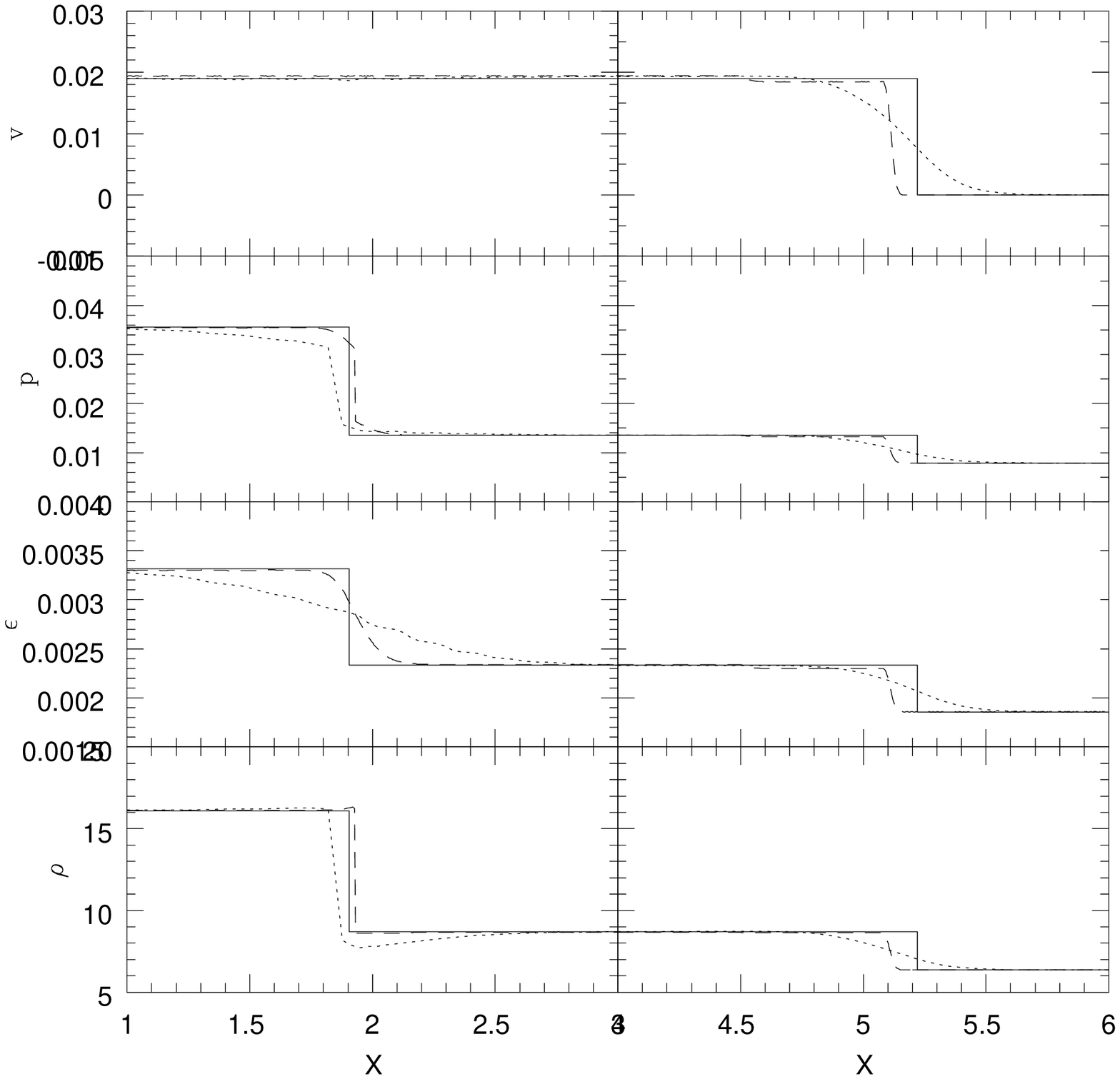}}
\caption{Enlargement of discontinuities of Fig. 1 both for $h = 5 \cdot 10^{-2}$ (long dashes) and for $h = 5 \cdot 10^{-3}$ (dots) SPH-Riemann (SPH+EoS) resolution lengths. The analytical solution is also plotted (solid line).}

\resizebox{\hsize}{!}{\includegraphics[clip=true]{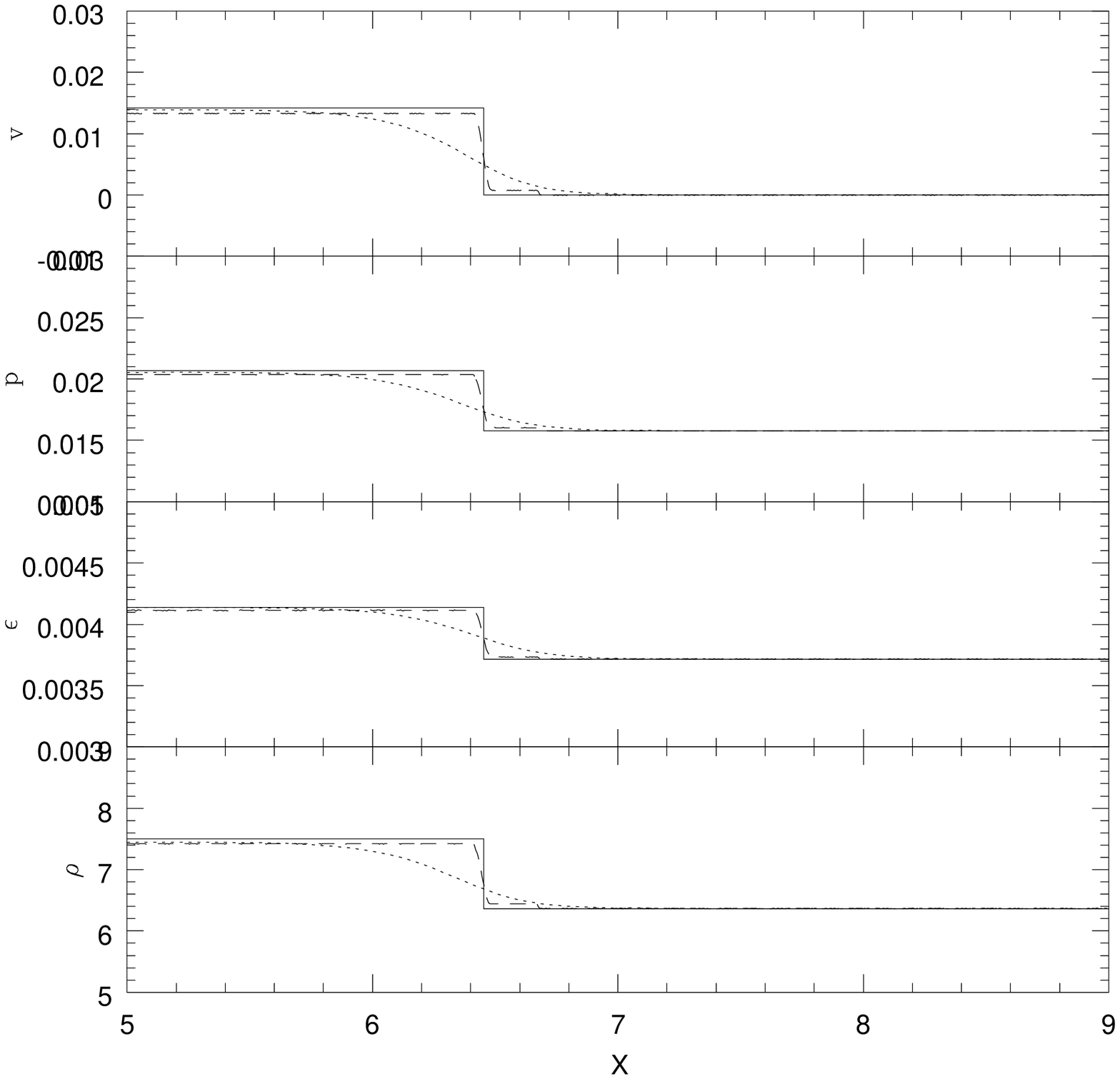}}
\caption{Enlargement of discontinuities of Fig. 2 both for $h = 5 \cdot 10^{-2}$ (long dashes) and for $h = 5 \cdot 10^{-3}$ (dots) SPH-Riemann (SPH+EoS) resolution lengths. The analytical solution is also plotted (solid line).}
\end{figure}

\begin{figure}
\resizebox{\hsize}{!}{\includegraphics[clip=true]{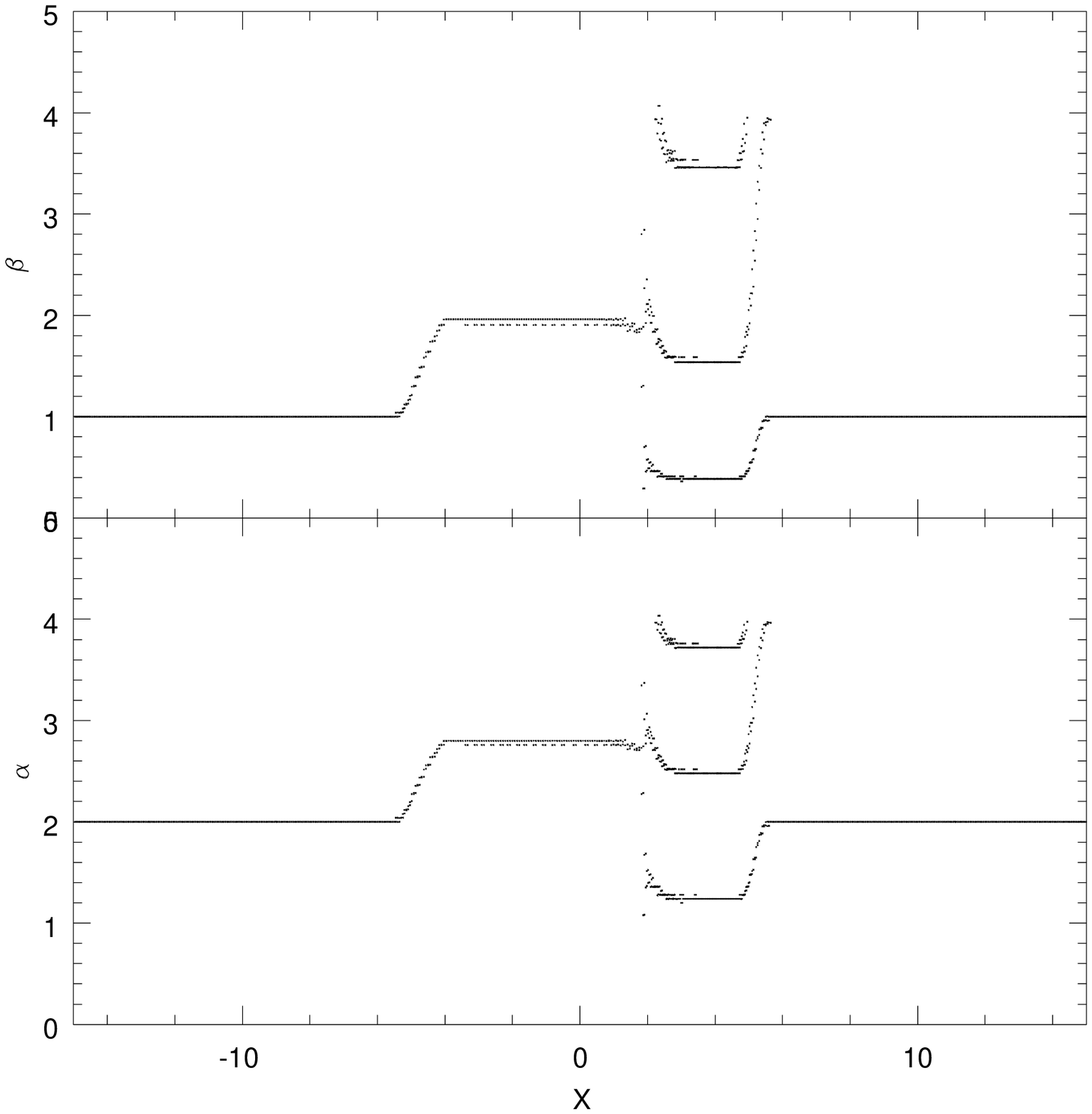}}
\caption{Parameters $\alpha$ and $\beta$ of the artificial viscosity counterpart for the SOD 1D shocktube test of Fig. 1.}

\resizebox{\hsize}{!}{\includegraphics[clip=true]{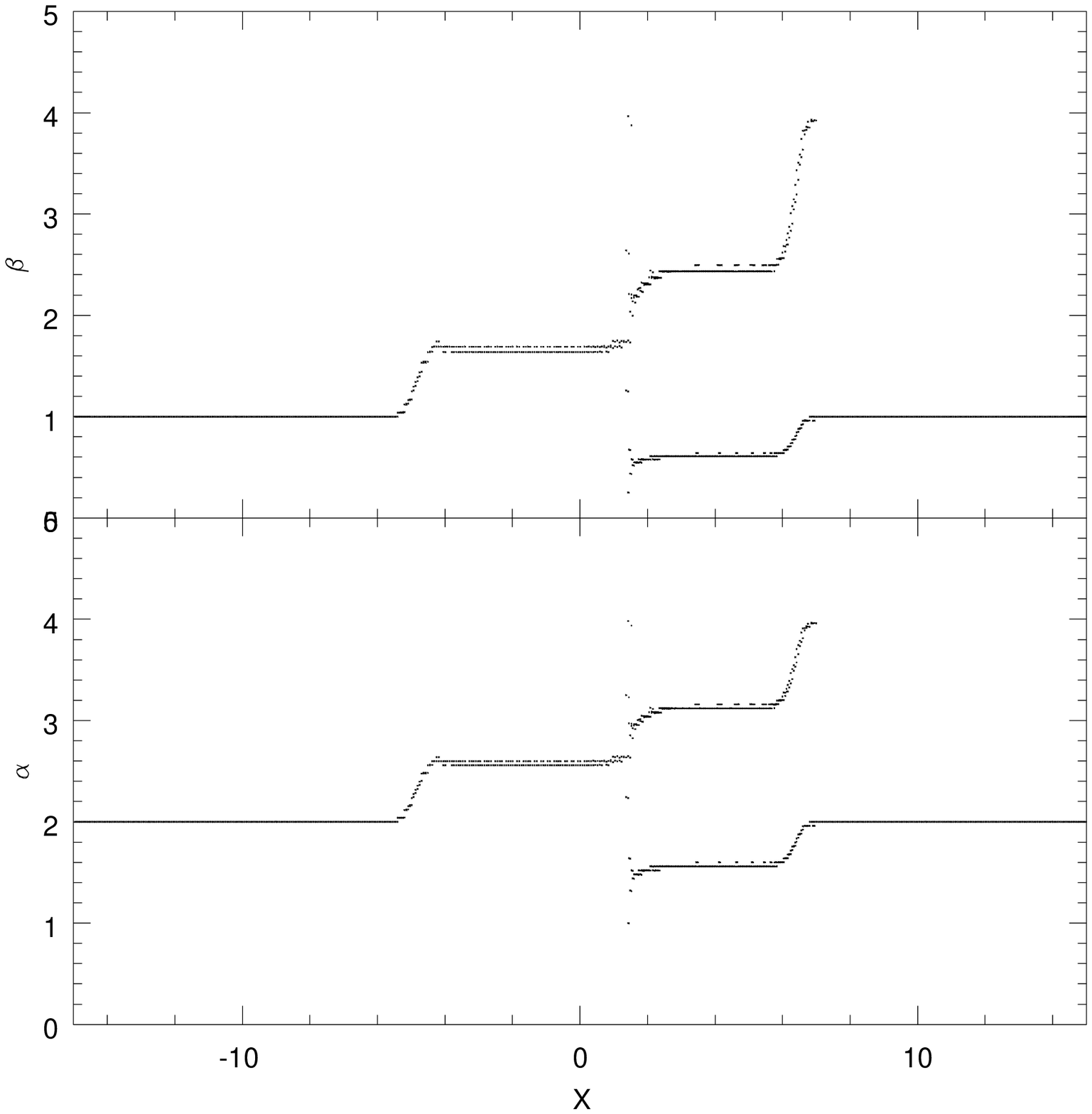}}
\caption{Parameters $\alpha$ and $\beta$ of the artificial viscosity counterpart for the SOD 1D shocktube test of Fig. 2.}
\end{figure}

  In this section a comparison of analytical and our SPH-Riemann (SPH+EoS) 1D shock tube test results \citep{a32}, also with the initial particle configuration (time $T = 0$), is made. Notice that the so called analytical solution of the Riemann problem is obtained through iterative procedures left-right to the discontinuity using the Rankine-Hugoniot "jump" solutions. Figg. 1 and 2 show results concerning the particle density, thermal energy per unit mass and velocity, after a considerable time evolution at time $T = 100$. The whole computational domain is built up with $2001$ particles from $X = 0$ to $X = 100$, whose mass is different, according to the shock initial position. At time $T = 0$ all particles are motionless. while the ratios $\rho_{1}/\rho_{2} = 3$ and $\epsilon_{1}/\epsilon_{2} = 2$ in Fig. 1, and  $\rho_{1}/\rho_{2} = 3$ and $\epsilon_{1}/\epsilon_{2} = 1$ in Fig. 2, between the two sides left-right. The first $5$ and the last $5$ particles of the 1D computational domain, keep fixed positions and do not move. The choice of the final computational time is totally arbitrary, since the shock progresses in time. $v = 0$ at the beginning of each simulation. Hence the adimensional temporal unity is chosen so that $\int_{0}^{l} dx/c_{s} = 1$. Being the sound velocity initially constant, this mathematically means that $\int_{0}^{l} dx = c_{s}$, so that $l = c_{s}$.

  Results, where our SPH-Riemann (SPH+EoS) solution is applied to SPH, display a good comparison with the analytical solution. Discrepancies involve only $4$ particle smoothing resolution lengths at most, except for numerical solutions corresponding to analytical vertical profiles regarding thermal energy where, as SPH solutions, discrepancies are larger. This means that, according to the cautionary remarks in \S 2.3, the physical dissipation introduced in the EoS (eqs. 36, or 37+38, or 56, or 60) is effective. Any arbitrary artificial viscosity contribution is totally absent, as well as any thermodynamic properties of neighbour particle are taken into account, as the application of Godunov-type solvers \citep{a18,a46,a21,a20,a45,a22} does. In SPH+EoS approach, a less evident Gibbs phenomenon \citep{a50,a51}, up to $20 \div 30 \%$ less affects the numerical solution.

  To check the reliability of the adopted EoS (eqs. 36, or 37+38, or 56, or 60), the same 1D shock tube tests were performed adopting a smaller particle smoothing resolution length $h = 5 \cdot 10^{-3}$, working with 20001 particles and adopting the same initial and boundary conditions. This check is fundamental to prove that particles do not interpenetrate with each other and to verify the final results as a function of the improved adopted spatial smoothing resolution. Results are shown in Figg. 3 and 4, where the comparison with both previous results and with the analytical solution are also reported. If results are good enough in a low-medium smoothing resolution, they are even better in higher spatial resolution, confirming that the shock profile depends on spatial resolution length. However this is a natural result emerging whichever is the adopted numerical technique.

  For the SPH+EoS modelling where eqs. (36, or 37+38, or 56) are used, the SPH $\alpha$ and $\beta$ parameters of artificial viscosity counterpart (eqs. 42, 43) are shown in Figg. 5 and 6 for both Sod shock tube tests. These calculations must be considered for each neighbour of each particle because they take into account both $c_{si}/c_{sij}$ and $|\bmath{r}|_{ij}/h$, as well as both $c_{si}^{2}/c_{sij}^{2}$ and $|\bmath{r}|_{ij}^{2}/h^{2}$. In the flat motionless zones, far from the shock discontinuities, $|\bmath{r}|_{ij} = h$, so that $\alpha = 2$ and $\beta = 1$, whilst those neighbours whose $|\bmath{r}|_{ij} = 2h$ are not taken into account because SPH interpolations are ineffective at the Kernel edge. In these motionless regions, despite $\alpha = 2$ and $\beta = 1$, any physical dissipation in SPH+EoS is not active, as in SPH, because $\bmath{r}_{ij} \cdot \bmath{v}_{ij} = 0$.  Notice that, according to the EoS eqs. (36, or 37+38, or 56), whenever $\bmath{r}_{ij} \cdot \bmath{v}_{ij} \geq 0$, any dissipation is physically prevented. Thus, whichever are the $\alpha$ and $\beta$ counterparts, their role is useless. Both $\alpha$ and $\beta$ assume multiple values wherever any kinetic evolution of the flow is recorded, especially in the proximity of the discontinuities, not only because of variations of particle mutual separation and sound velocity, but also because of small statistical fluctuations in the particle kinematic and thermal properties. This shows both the tendency for the closer interacting particles (smaller $|\bmath{r}|_{ij}/h < 1$) to have smaller $\alpha$ and $\beta$ values - whilst it is the opposite for the farther interacting particles (larger $1 < |\bmath{r}|_{ij}/h < 2$), in so far as $\bmath{r}_{ij} \cdot \bmath{v}_{ij} < 0$ - and very small horizontal "dual" behaviour of $\alpha$ and $\beta$ values. Besides, as Figg. 5 and 6 show, any  smoothing on the vertical discontinuities reflects the SPH smoothing of vertical linear tracks in the shock profiles.

\begin{figure*}
\resizebox{\hsize}{!}{\includegraphics[clip=true]{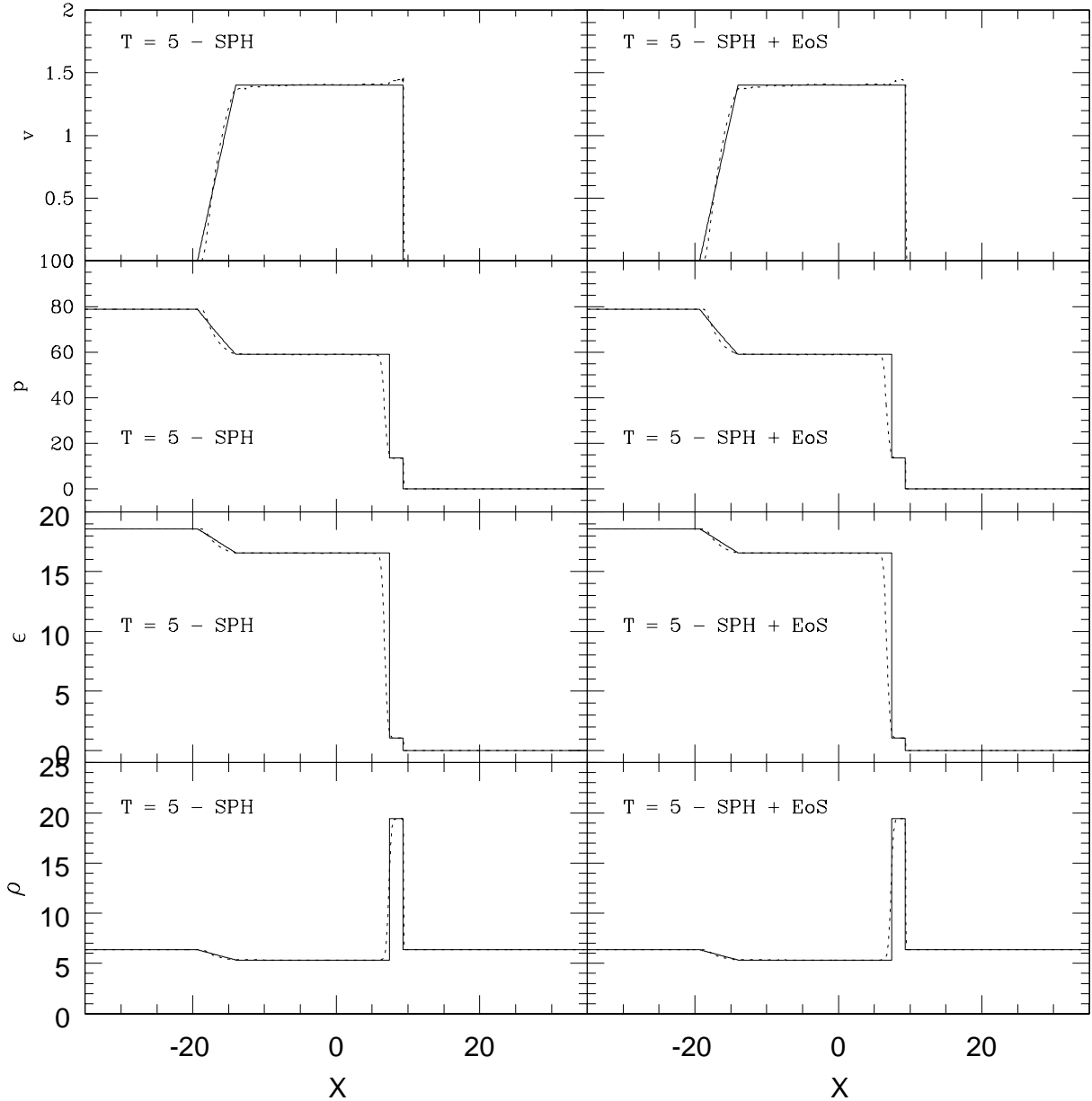}}
\caption{1D blast wave test as far as both analytical (solid line) and our SPH-Riemann (SPH+EoS, dots) results are concerned (right side plots). Density $\rho$, thermal energy $\epsilon$, pressure $p$ and velocity $v$ are plotted at time $T = 5$. Density and thermal energy of particles initially at rest at time $T = 0$ refer to values plotted at the two edges of each plot. The initial velocity is zero throughout. SPH results are also reported (left side plots).}
\end{figure*}

\begin{figure}
\resizebox{\hsize}{!}{\includegraphics[clip=true]{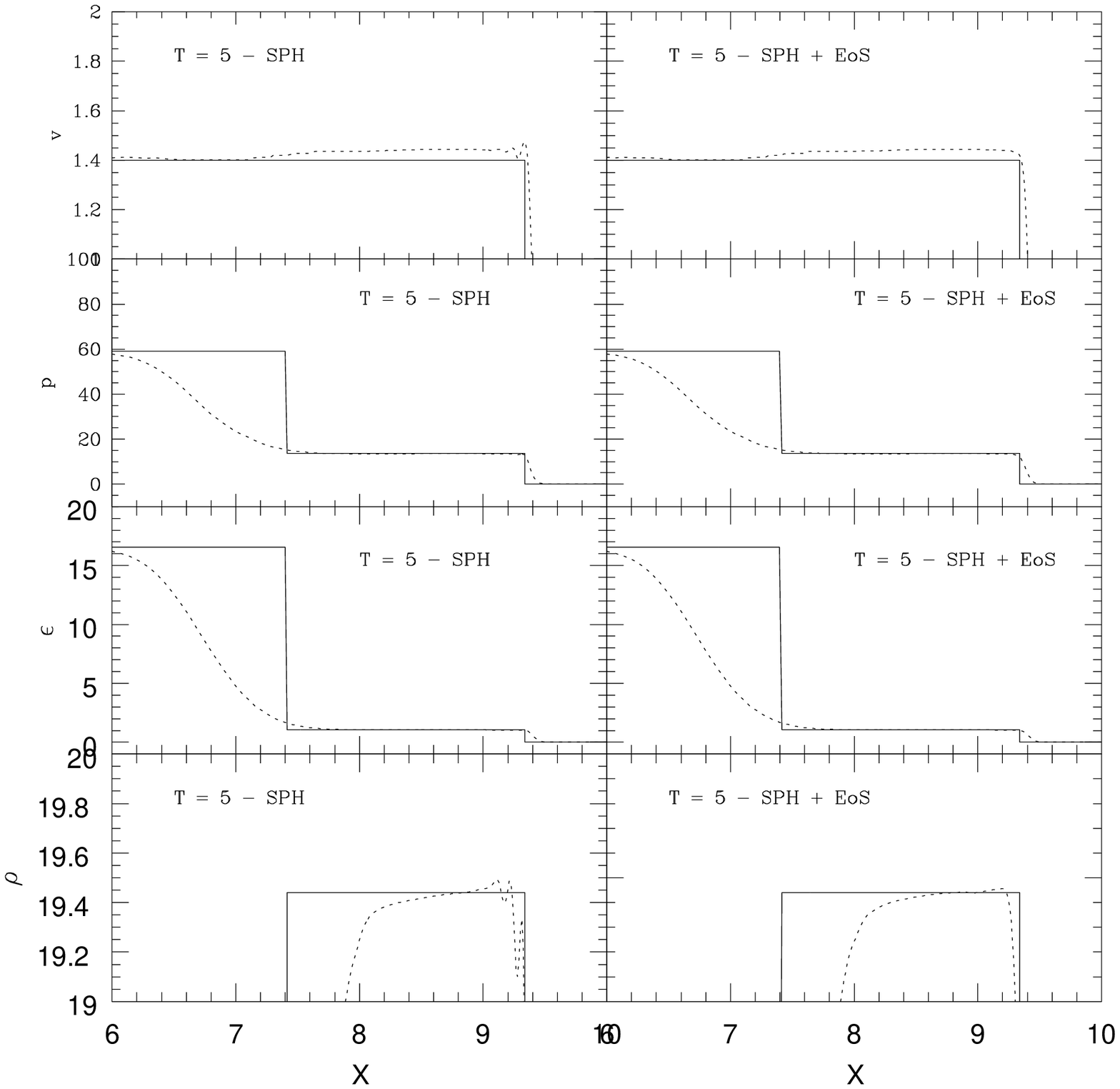}}
\caption{Enlargement of previous Fig. 7 showing SPH instabilities at contact discontinuity. The Eos+SPH profile (both in dots) does not of such instabilities. The analytical profile (solid line) is also shown.}

\resizebox{\hsize}{!}{\includegraphics[clip=true]{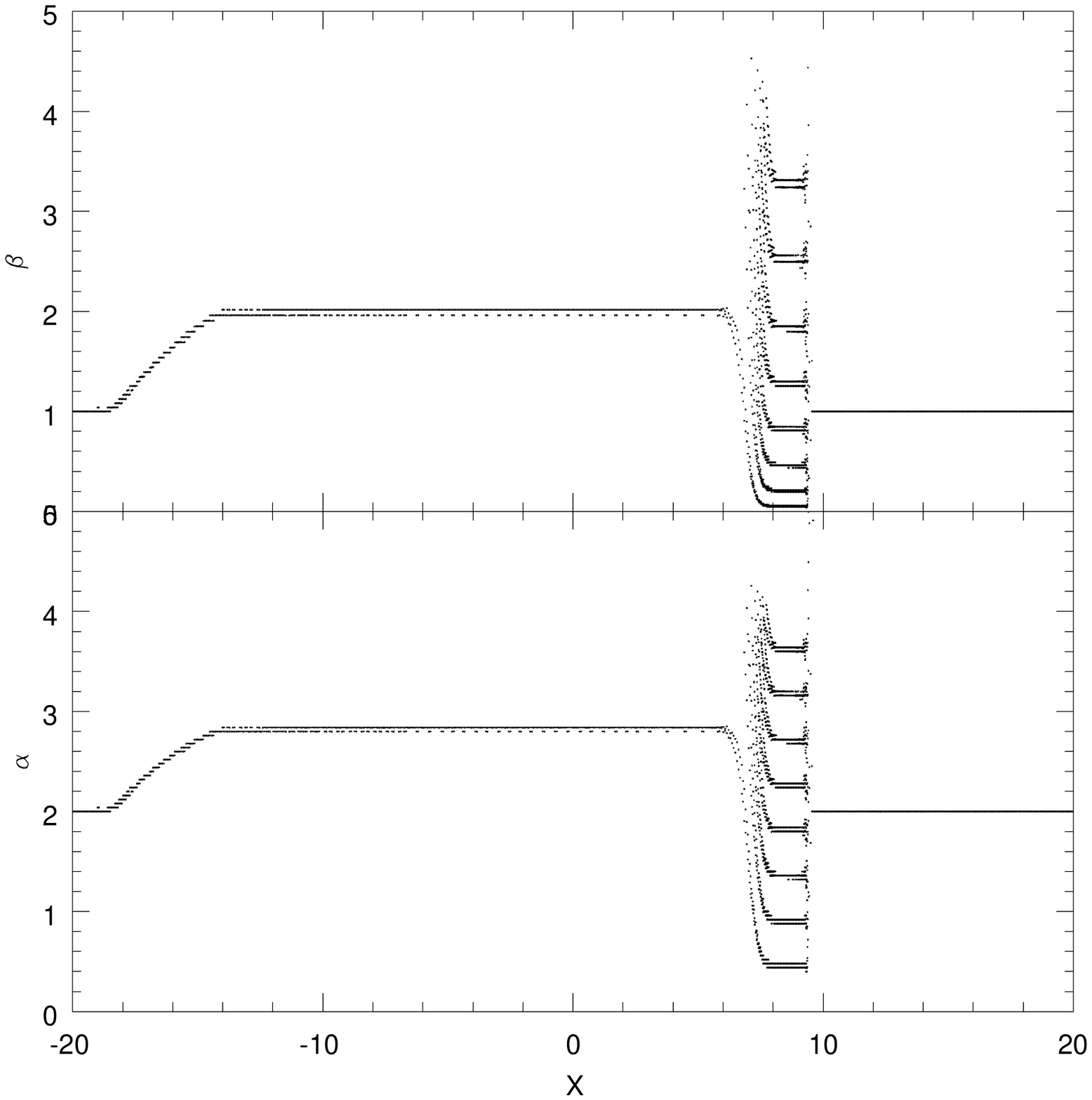}}
\caption{Parameters $\alpha$ and $\beta$ of the artificial viscosity counterpart for the 1D blast wave test of Fig. 8.}
\end{figure}

\subsection{1D Blast wave}

  Whenever in a shocktube the ratios $p_{1}/p_{1} = \epsilon_{1}/\epsilon_{2} \gg 1$, and consequently $\rho_{1}/\rho_{2} = 1$, and $v_{1} = v_{2} = 0$, such a discontinuity is called a "blast wave". Being $v = 0$ at the beginning of each simulation, the adimensional temporal unity is chosen as previously written in the 1D Sod shocktube test before. Figg. 7 and 8 show a comparison of SPH and SPH+EoS results with the so called analytical solution, as shortly described in the previous subsection, after a considerable time evolution at time $T = 5$. However, especially in these cases, the analytical solution is considered corrected in so far as $\rho_{1}/\rho_{2} \leq (\gamma + 1)/(\gamma - 1)$. In the blast wave test here considered, $p_{1}/p_{1} = \epsilon_{1}/\epsilon_{2} = 10^{4}$, while other spatial, initial and boundary conditions, as well as the particle spatial smoothing resolution length are identical to those chosen in the previous test. Figg. 7 and 8 show that SPH and SPH+EoS results globally compare with each other and that they also compare with the analytical solution wherever $\rho_{1}/\rho_{2} \leq (\gamma + 1)/(\gamma - 1)$, that is wherever the Rankine-Hugoniot jump conditions hold. Beyond this limit, even the so called analytical solution is considered incorrect. Being $\gamma = 5/3$, the comparison is meaningful within $\rho_{1}/\rho_{2} \leq 4$. However, the SPH+EoS modelling has the advantage of offering a better solution, compared to the SPH one, for $\rho_{1}/\rho_{2} > 4$ (and therefore for $\rho_{1}/\rho_{2} > (\gamma + 1)/(\gamma - 1)$) as Fig. 8 clearly shows, where the SPH numerical solution suffers from some instabilities. Moreover, the SPH+EoS solution does not seem to suffer from any "blimp" effect at the contact discontinuity, as briefly discussed in \citet{a17}, where a modification of the artificial viscosity term is proposed (see also \S 2.2), as far as the velocity profile is concerned. Such effect comes out whenever a spatial high smoothing resolution is working together with an explicit handling of dissipation through an artificial viscosity damping to solve the Riemann problem of flow discontinuities. A low spatial smoothing resolution hides this effect because of the higher artificial damping due to a higher particle smoothing resolution length $h$ (eq. 15-16). Moreover, in SPH, even the choice of the arbitrary parameters $\alpha$ and $\beta$ should be linked to the specific physical problem. Instead, in SPH+EoS, the damping is strictly physical and local and any "blimp" effect is strongly reduced, especially for 1D blast waves, where strong discontinuities in the flow deeply affect the SPH numerical solution producing intrinsic numerical instabilities close to the propagating discontinuities (Figg. 7-8). The higher the spatial smoothing resolution (the smaller $h$), the higher the "blimp" instabilities.

  Fig. 9 shows the $\alpha$ and $\beta$ parameters of the artificial viscosity counterpart for the 1D blast wave of Fig. 7. Computational considerations for both parameters are equivalent to those written in the previous 1D Sod shock tube test. Notice that in such a test, in the denser region of the blast wave, the number of neighbour particles involved in the computation of such parameters is larger than in the previous test, as evidenced by the multiple $\alpha$ and $\beta$ values.

\subsection{2D radial spread and migration of an annulus Keplerian ring}

  The 2D radial spread and migration of an annulus ring is widely described in \citet{a85} in the case of a constant physical viscosity $\nu$. At time $T = 0$, the surface density, as a function of the radial distance $r$, is described by a Dirac $\delta$ function: $\Sigma (r,0) = M \delta (r - r_{\circ})/2 \pi r_{\circ}$, where $M$ is the mass of the entire ring and $r_{\circ}$ is its initial radius. The following time, the surface density is computed via standard methods as a function of the modified Bessel function $I_{1/4} (z)$:

\begin{equation}
\Sigma (x, \tau) = \frac{M}{\pi r_{\circ}^{2}} \tau^{-1} x^{-1/4} e^{- \frac{1 + x^{2}}{\tau}} I_{1/4} (2x/\tau),
\end{equation}

where $x = r/r_{\circ}$ and $\tau = 12 \nu T/r_{\circ}^{-2}$, expressing radial distances in $r_{\circ}$ units and time in viscous dissipation time units, respectively. $\int_{S} \Sigma (x, \tau) dS = 2 \pi \int \Sigma (x, \tau) dr =$ const equals the annulus mass throughout. Time is normalized so that $T = 1$ corresponds to the Keplerian period relative to the ring at $r_{\circ} = 1$. Examples of SPH spread on this argument can be found in \citet{a86,a89,a87}, as well as in \citet{a88} in SPH physically inviscid hydrodynamics on the basis that the SPH shear dissipation in Eulerian non viscous flows can be compared to physical dissipation \citep{a1,a9,a11} in a full Navier-Stokes approach. In particular an exhaustive comparison can also be found in \citet{a43,a44}.

  A significant comparison of SPH+EoS (eqs. 36, 37+38, or 56, not eq. 60) to SPH is shown in Fig. 10, where $XY$ density contour map plots are shown at the same times, to show whether dissipation in SPH+EoS approach gives results better fitting to the analytical viscous solution ($\nu \approx c_{s} h$) than the classical SPH dissipation. The radial distributions of surface density are shown in Figg. 11-12, as well as the radial profile of the theoretical surface density, according to the restricted hypotheses of the standard mechanism of physical dissipation (constant dissipation, zero initial thickness). As in \citet{a89,a87}, the initial ring radius is at $r_{\circ} = 1$, whose thickness is $\Delta r =0.5$, is composed of $40000$ equal mass ($m_{i} = 2.5 \cdot 10^{-15}$) pressureless Keplerian ($\bmath{v} = \bmath{v}_{Kepl}$, $\nabla \cdot \bmath{v} = 0$ at $T = 0$) SPH particles, with $h = 9 \cdot 10^{-2}$, with $c_{s} = 5 \cdot 10^{-2}$, and with initial density radial distribution corresponding to the analytical solution at time $T$, whose $\tau = 0.018$. To this purpose, a random number generator has been used, as in \citet{a87}. The central accretor mass is normalized to $M = 1$. The kinematic shear dissipation is estimated as $\nu \propto c_{s} h$ \citep{a1,a9,a11,a43,a44}. SPH results of this test compare with those of \citet{a89,a87} and of \citet{a88}. Being in an Eulerian fluid dynamics, in SPH the shear effect is produced by the artificial viscosity, and the consequent artificial pressure terms, having removed the physical pressure terms. \citet{a87} discussed its limits, treating a ring free of any pressure force, where a viscous stress tensor similarity of viscosity is considered. In SPH+EoS the physical dissipation, appearing as further terms in the EoS, determines the shearing effects, which is still effective also removing the physical pressure terms as in the SPH counterpart. It is decisive the fact that the SPH+EoS radial density much better fit both the theoretical radial profile and its radial migration towards the central accretor. Notice that these results are obtained according to the cubic spline (8) Kernel analytical expression and according to the correlation $\nu \sim c_{s} h$ \citep{a1} in the expression where $\tau = 12 \nu T/r_{\circ}^{-2}$.

\begin{figure}
\resizebox{\hsize}{!}{\includegraphics[clip=true]{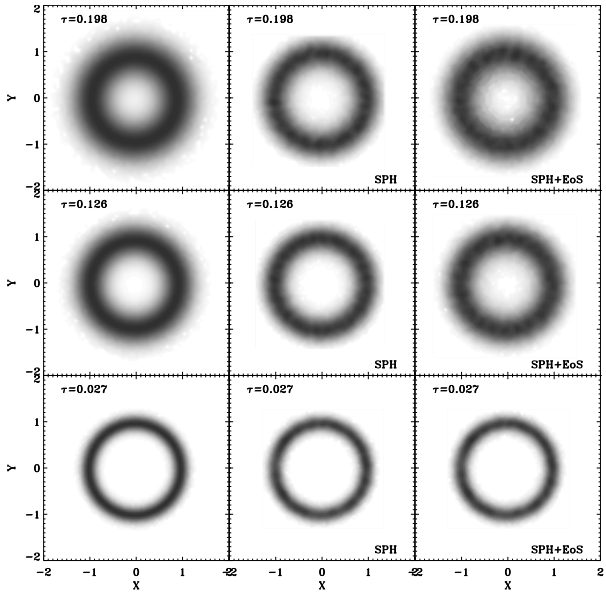}}
\caption{$XY$ plots of ring density contour maps. Times are reported for each configuration both theoretical and numerical (SPH or SPH+EoS).}
\end{figure}

\begin{figure}
\resizebox{\hsize}{!}{\includegraphics[clip=true]{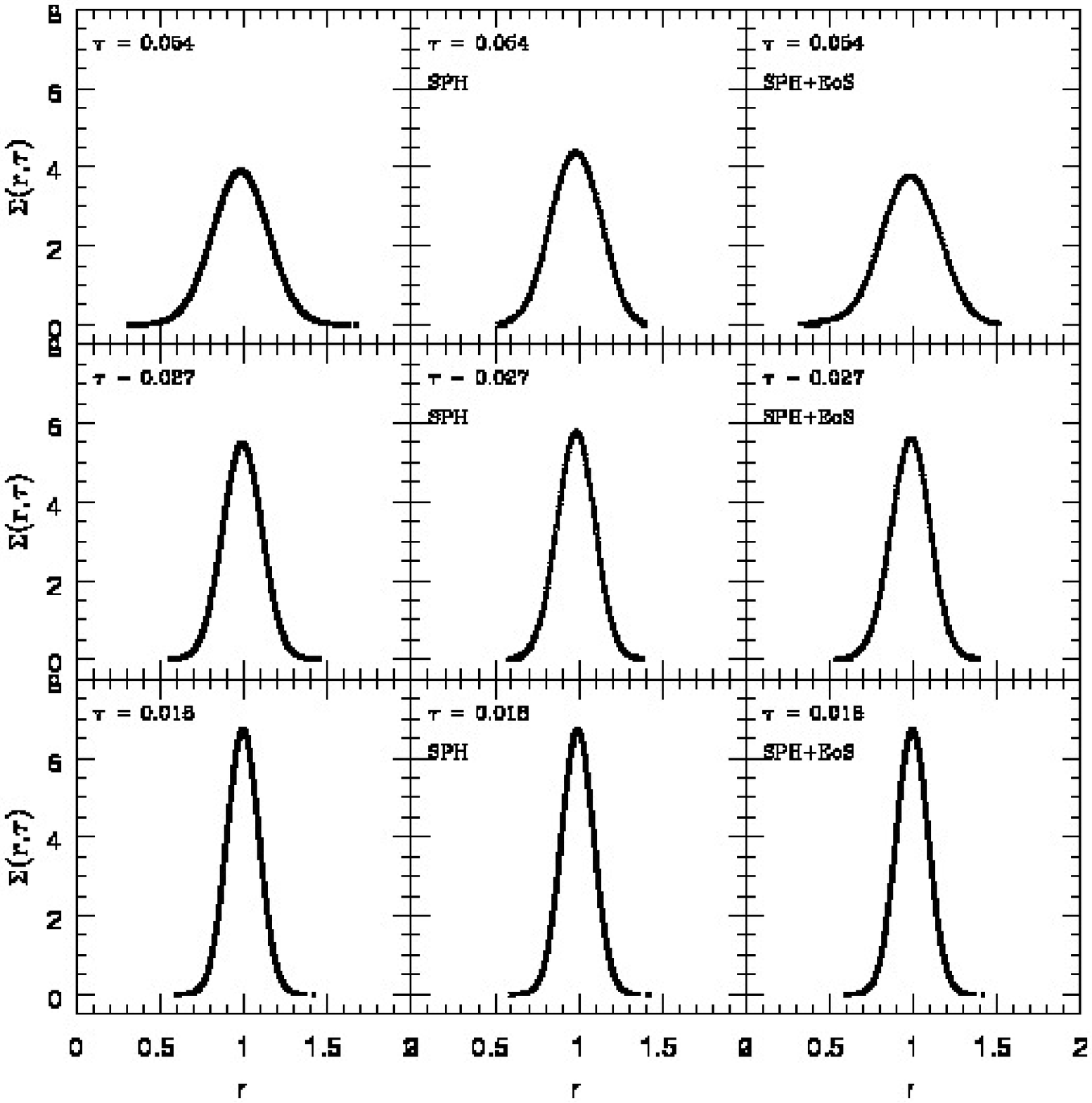}}
\caption{Surface density $\Sigma (r, \tau)$ in $10^{-11}$ units as a function of radial distance from initial configurations at $\tau = 0.018$ when radial profile equals the theoretical analytical one. Subsequent snapshots are reported for each configuration both theoretical (left side plots) and numerical (SPH or SPH+EoS).}

\resizebox{\hsize}{!}{\includegraphics[clip=true]{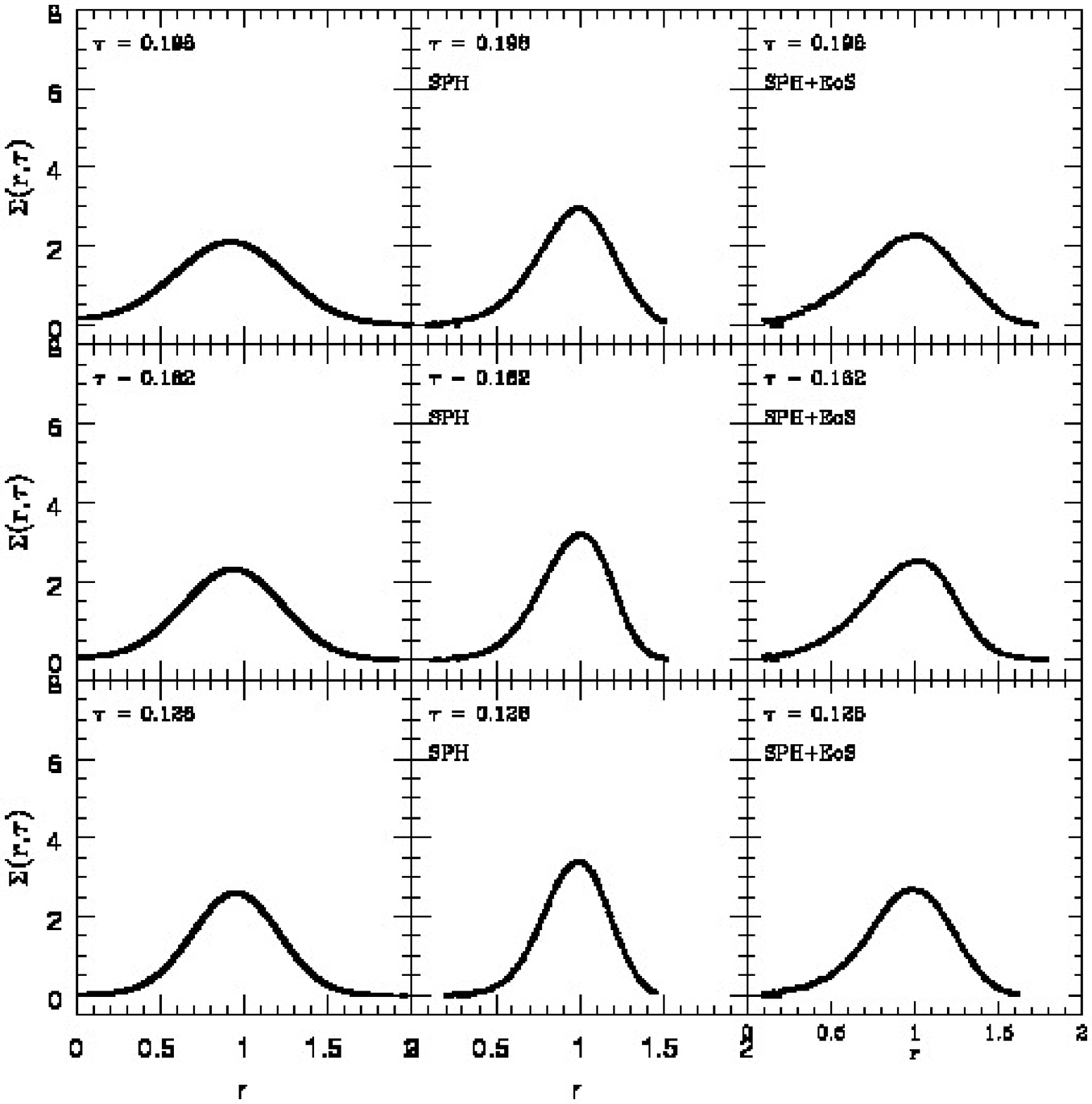}}
\caption{As Fig. 11 for more evolved times.}
\end{figure}

  The arbitrary assumption of a different $\nu$ would involve the appropriate time $T$ to get the same $\tau$. In fact, the variation of the initial $h$, and/or $c_{s}$, to get a smaller (larger) $\nu$ implies a longer (shorter) time $T$ to get the same $tau$ to produce the same radial distribution of particles.

  Instead, different radial distributions are obtained assuming a different interpretation of the analytical expression for $\nu$. In fact, considering $\nu \sim 0.1 \alpha_{SPH} c_{s} h$ \citep{a9,a11}, with $\alpha_{SPH} \approx 1$, it is necessary to perform the numerical simulations for a time $T$ ten times longer to get the same $\tau$, in spite of the initial $h$ and $c_{s}$ are unchanged. This involves a too large annulus spread as far as the numerical results are concerned in both cases, compared to the analytical solution. To this purpose, even the successful test in \citet{a92,a9} is not conclusive because in those papers the comparison with the analytical solution is successful up to a time $T = 0.8$ which, in that parametrization, would involve a too short final $\tau \sim 3.32 \cdot 10^{-4}$, that is much shorter than our $\tau$ scales. This is also demonstrated in their results because, together with their annulus spread, a very short migration of the entire annular distribution towards $r = 0$ is also shown as a consequence of the fact that the final ring configuration is due to a too short time evolution. Instead, SPH results of \citet{a89,a87,a88} compare with our SPH results because the spread and migration is not meaningless. It is the SPH+EoS that is better in so far as $\nu \sim c_{s} h$ is considered.
  
  $\alpha$ and $\beta$ artificial viscosity parameters of the SPH counterpart, for SPH+EoS modelling according to eqs. (42, 43), are simply expressed by $\alpha = 2 r_{ij}/h$ and $\beta = (r_{ij}/h)^{2}$, since the whole annulus ring is isothermal in such shockless and pressureless simulations. This means that, in their radial profile - here not shown for the sake of simplicity - both parameters span from small positive values in the denser regions of the annulus ring, up to their maximum values ($\alpha = \beta = 4$), at the beginning of each SPH+EoS model. However, their minimum values increase, towards their maximum limit ($4$), because of the increase of the particle mean free path, as time goes by.

  Circular rings, appearing in Fig. 10 for both numerical schemes, are an unavoidable effect due to the Lagrangian particle-based technique, as discussed in \citet{a89,a87}. This effect cannot be present throughout in the theoretical $XY$ plots because of the random-based representation of data points.
  
  Instead, if the strictly annulus ring has to be strictly Keplerian, being $\nabla \cdot \bmath{v} = 0$, the SPH+EoS form (60) does not produce any radial transport, being zero the intrinsic physical dissipation. Hence, the same ring configuration at $\tau = 0.018$ stays endlessly the same. Hence, results on the radial spread and radial migration of such a test are not shown.

\subsection{2D slipping flow of a single row of particles within a horizontal bounded pipe}

\begin{figure}
\resizebox{\hsize}{!}{\includegraphics[clip=true]{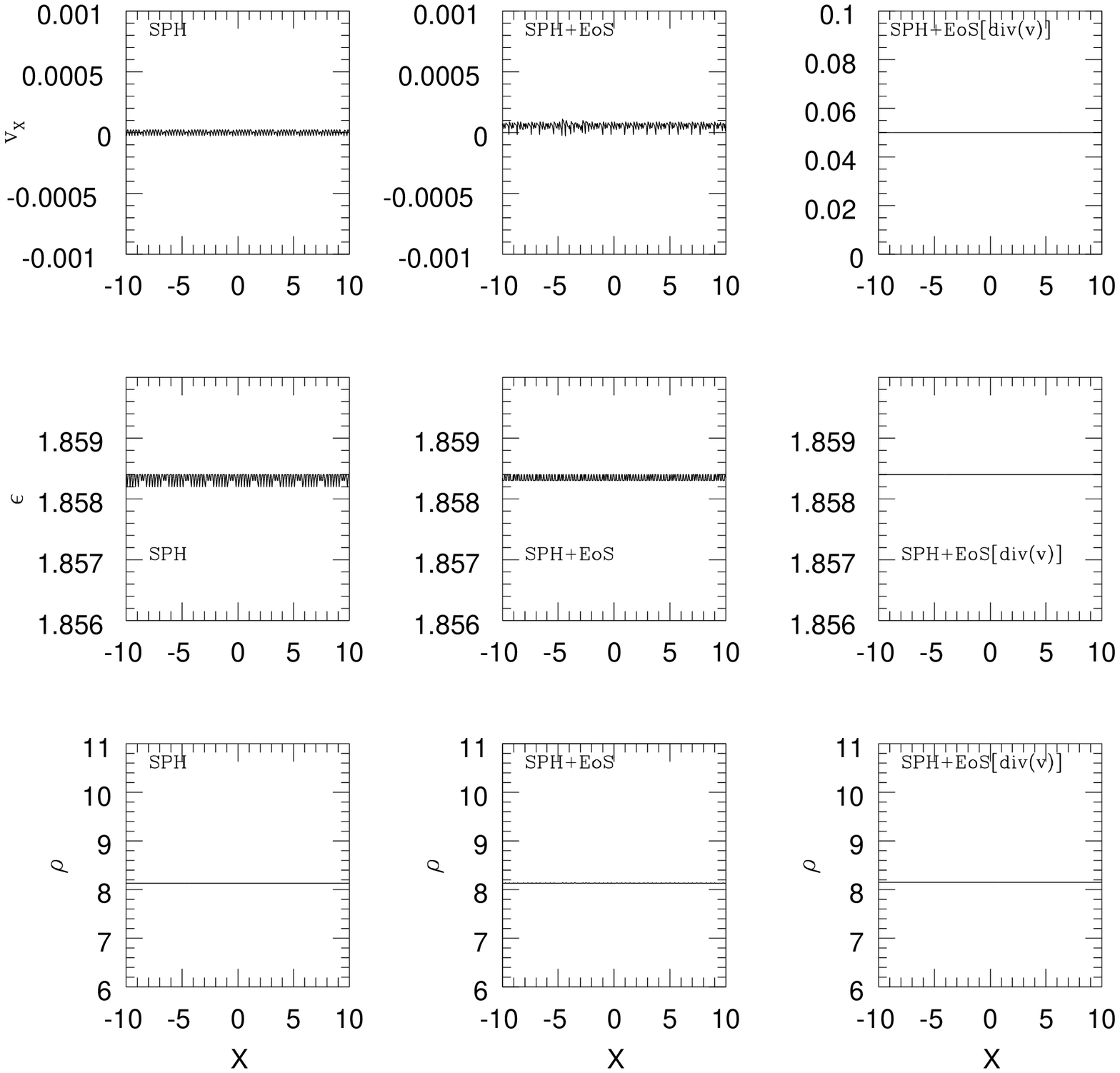}}
\caption{Particle density, thermal energy per unit mass (in $10^{3}$ units) and velocity for SPH and SPH+EoS modelling. Time $T = 100$. SPH+EoS[div(v)] refers to SPH+EoS modelling where eq. (60) is applied.}
\end{figure}

  In this test we compare results concerning the ideal flow of one horizontal row of particles in a horizontal bounded pipe. To this purpose, nine horizontal rows of equal mass particles are considered, whose $m = 10^{-3}$. The computation is a fully 2D computation. The four up and the four down external rows are densely populated of permanently isothermal fixed particles, whose mutual horizonthal separation equals $h/4$, whose thermal energy per unit mass equals that of the flowing particles within the pipe. The intermediate flowing row of particles consists of equally spaced particles whose mutual separation equals $h$ and whose uniform velocity $v = h = 5 \cdot 10^{-2}$. The initial Mach number is $\sim 3.5 \cdot 10^{-2}$ throughout. $h/4$ is also the mutual vertical separation throughout. Our purpose is to determine a permanent condition where $\nabla \cdot \bmath{v} = 0$ because the analytical solution for an ideal gas is that of a constant flow, whose velocity and whose thermodynamic properties are constant in time. To this purpose, the denser are the boundaries (in terms of numerical density), the better are such conditions since $\nabla \cdot \bmath{v}$ statistically better equals zero. In such a test, the analytical solution is perfectly known. However, our purpose is to verify whether SPH or SPH+EoS in its (36, or 37+38, or 56), or in its (60) formalism (hereinafter SPH+EoS[div(v)]) compare with each other and/or compare with the analytical solution, taking into account that either the artificial viscosity in SPH or the physical dissipation in the SPH+EoS formalism (eqs. 36, or 37+38, or 56) activate whenever two particles approach each other. At the same time we also check whether the SPH+EoS[div(v)] in its form (eq. 60) correctly behaves because its physical dissipation depends on $\nabla \cdot \bmath{v}$. Whichever is the final computational time, both $\bmath{v}$ and the thermal energy per unit mass $\epsilon$ for each particle should be constant. Any deviation from their initial values is an evaluation of the inadequacy of the numerical approach to the true solution. Of course, we pay attention that, according to eqs. (42) and (43), the physical dissipation in the SPH+EoS formalism, according to eqs. (36, or 37+38, or 56, not 60), also depends on mutual particle separation and, to this purpose, the shorter is the vertical separation among particle rows, the smaller is the physical dissipation according to the Riemann problem solution, up to its natural limit towards the 1D Riemann problem solution.

  Fig. 13 shows results of such simulations at time $T = 100$ concerning particle density, thermal energy per unit mass and velocity (subsonic in such an example, with $\gamma = 5/3$ and $\epsilon \approx 1.858 \cdot 10^{3}$). Both SPH and, especially SPH+EoS (referred to eqs. 36, or 37+38, or 56, not 60), show statistical noise around $v = 0$, due to the fact that dissipation - artificial, or physical - activates at the particle approaching and in particular among the flow particle in the pipe with the two tight boundaries, preventing the correct uniform slipping of the particle flow, since from the beginning of simulation. Instead, results for the SPH+EoS case where the EoS takes into account of $\nabla \cdot \bmath{v}$ (eq. 60), perfectly fit the analytical solution without any  - or very small - vibrations in the velocity field, and consequently on $\epsilon$. Notice that such vibrations in the velocity field are a bit larger when SPH+EoS is considered according to eqs. (36, or 37+38, or 56, not 60). This is mainly due to the fact that the $\alpha$ and $\beta$ counterpart are statistically larger than $\alpha = 1$ and $\alpha = 1 \div 2$, as currently adopted in SPH for the solution of the Riemann problem. Density, as well as the thermal energy per unit mass are statistically not perturbed because of $\nabla \cdot \bmath{v} \approx 0$ even where $v \approx 0$ both in SPH and in SPH+EoS approach.

\section{Discussion and conclusions}

  The comparison of 1D Sod shock tube analytical results with SPH-Riemann (SPH+EoS) ones, where a revision of the EoS within the Riemann problem is made, are quite successful. In particular, in our modelling, neither a parametrized artificial viscosity, nor any dependence on spatial smoothing resolution length $h$, nor a sophisticated Godunov solver, nor additional computational time are involved.

  The simple EoS for inviscid ideal gases: $p = (\gamma - 1) \rho \epsilon$ cannot be strictly applied in shock problems because of the fact that, solving the Euler equations, not only instabilities and spurious heating come out, but also that this EoS derived by either the Charles, Volta Gay-Lussac and Boyle-Mariotte laws or by a partition function of statistical thermodynamics, should be strictly applied only either in equilibrium configurations or to "reversible or quasi-static" evolutions of thermodynamic systems without any dissipation. This is a restriction that does not match with shock flow dynamics, when dissipation on the shock front cannot be neglected. Those techniques involving sophisticated Godunov-type schemes also introduce some useful dissipation mechanism in the numerical scheme \citep{a19}, right for obtaining correct solutions of Euler equations.

  To conclude, the general EoS here proposed, shows the correct behaviour, even in the presence of dissipative shocks in non viscous flows. A successful check of the reliability of the Riemann approach, here proposed on inviscid hydrodynamics, is shown in App. A, where a study is shown on the coming out of spiral structures and of shocks in the radial flow of accretion discs. The comparison, according to the different techniques, where artificial or physical dissipation is explicitly introduced, shows that the entire disc structure and dynamics, as well as the coming out of spirals \citep{a34,a36,a37,a35,a38}, are much better evidenced in those simulations where the EoS and the related dissipation are treated in their full physical sense according to the Riemann problem solution, where the dissipation is the most effective, while the EoS (eq. 60), where physical dissipation is correlated to the flow compression ($\nabla \cdot \bmath{v}$), shows the lower shear dissipation, being the particle tangential kinematics in the disk bulk more comparable to the Keplerian velocity field.

\subsection{Concluding remarks}

\begin{itemize}
  \item The need to introduce a numerical dissipation (implicit or explicit) is necessary to solve the hyperbolic Euler equation system in non viscous flow dynamics. However, there is also a physical motivation because shock phenomena have to be considered as irreversible events.

  \item Whichever is the adopted numerical technique for the non viscous hydrodynamics, a problem arises as far as the choice of some either explicit or implicit (adopting a Godunov-like technique) free parameters for dissipation is concerned. As an example, we recall the choice for $\alpha$ and $\beta$ artificial viscosity coefficients in SPH, $\alpha^{\ast}$, $\beta$ and $S_{i}$ in the formulation of \citet{a16}, the calculation of pressure $p^{\ast}$ according to \citep{a17}, as well as whether either a thermal energy diffusion term or an attenuator \citet{a15} or both have to be used.

  \item For these reasons, even if working in SPH, we propose an EoS for non viscous ideal gases that:
  
  \subitem introduce a physical dissipation both for the resolution of the Riemann problem - to solve shocks - and for the resolution of shear flows;

  \subitem such a dissipation does not depend on arbitrary parameters, as well as on spatial smoothing resolution length $h$. It is correlated to $\alpha$ and $\beta$ through eqs. (42, 43, strictly valid for the Riemann problem), or through eqs. (62, 63 for a more general thematics), according to local thermodynamic conditions;

  \subitem it is justified according to statistical thermodynamics calculations;
  
  \item Analytic solutions for inviscid shocks and blast waves are highly idealized and subjected to Rankine-Hugoniot left-right ($l$, $r$) "jump conditions" limits ($\rho_{l}/\rho_{r} \leq (\gamma + 1)/(\gamma - 1)$). On theses conditions, analytical profiles are always "ruler-drawn". If shocks are irreversible thermodynamic events, a physical dissipation should smooth out every vertical linear profile of solutions;

  \item In this work, 1D profile of shocks and blast waves, where the physical dissipation is introduced in the EoS, better compare with the analytical solution and do not suffer from instabilities ("blimp") in the "flat" zones close to vertical discontinuities in the physical parameters. Vertical descending profiles also better compare. Zones where rarefaction waves exist compare with SPH solutions. A larger discrepancy is evident only in the thermal energy profiles wherever vertical ascending profiles occur. However, we recall that thermal dissipation is not applied;

  \item The equivalent formulations for the EoS of ideal gases (36, or 37+38, 56) appear successful for the treatment of the Riemann problem. Eq. (60) is also fair for non viscous shear flows. Nevertheless, even in not refined form (36, or 37+38 or 56), the physical dissipation, introduced in the EoS, better fits to the analytical solution of viscous flows if simply $\nu \approx c_{s} h$ \citep{a1} describes the SPH dissipation (\S 6.3). The EoS in the form of eq. (60) introduces a physical dissipation only when a real local gas compression occurs. Hence, for accretion problems, if the velocity field is mainly Keplerian, either a real physical turbulent viscosity \citep{a86,a49,a43,a44} in the Navier-Stokes approach, or an EoS in the Euler approach in the form of eqs. (36, or 37+38, or 56) should be considered;

  \subitem $\alpha$ and $\beta$ (eqs. 42, 43) counterpart of the SPH+EoS in the form of eqs. (36, or 37+38, or 56) are generally larger than those values largely adopted in SPH techniques $\sim 1$. This lets us think that $\alpha = 1$ and $\beta = 1 \div 2$ are a compromise, allowing description of both a Riemann problem case, as well as a shear flow at the same time, with the price of including of some small shortcomings. The same $\alpha$ and $\beta$ values are mostly close to $0$ for those flows where no gas compression occurs, being $\nabla \cdot \bmath{v} \geq 0$, when the SPH+EoS[div(v)] for of the Eos is taken into account in its more general (eq. 60) form. However in some cases, wherever and whenever a high density increase occurs: $\rho^{-1} d\rho/dt = - \nabla \cdot \bmath{v} \gg 0$, both parameters could reach considerable values.
\end{itemize}

\appendix

\section{The appearance of spirals in accretion discs in close binaries}

\begin{figure}
\resizebox{\hsize}{!}{\includegraphics[clip=true]{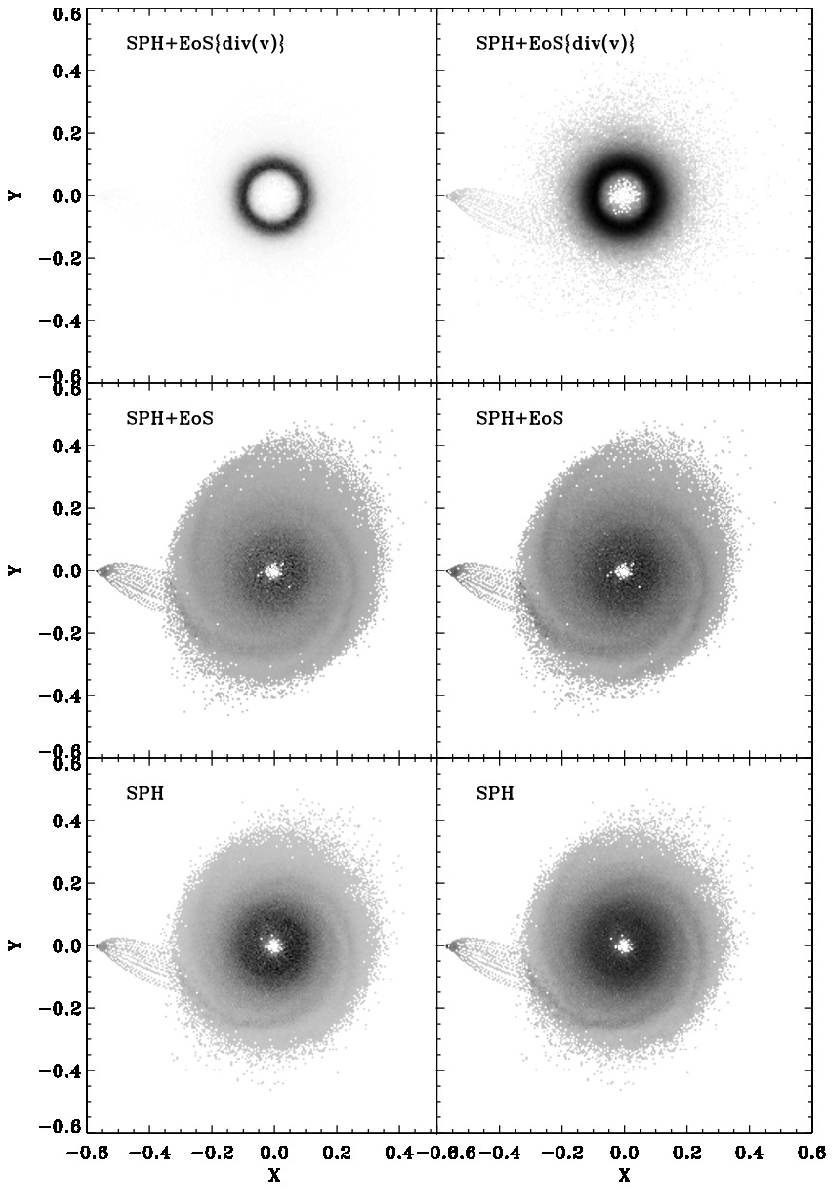}}
\caption{XY plots of $64$ greytones density $\rho$ isocontours of 3D disc modelling in CBs. A linear scale of greytones is represented on the left side, while its logarithmic scale is represented on the right. SPH refers to normal SPH results; SPH+EoS refers to SPH results where a reformulation of the EoS according to the Rieman problem solution is proposed; SPH+EoS[div(v)] refers to SPH results where a reformulation of the EoS in its more general physical sense is here proposed. When in statistically steady state, particles are $\sim 79696$ for SPH, $\sim 72023$ for SPH+EoS ans $\sim 112107$ for SPH+EoS[div(v)].}

\resizebox{\hsize}{!}{\includegraphics[clip=true]{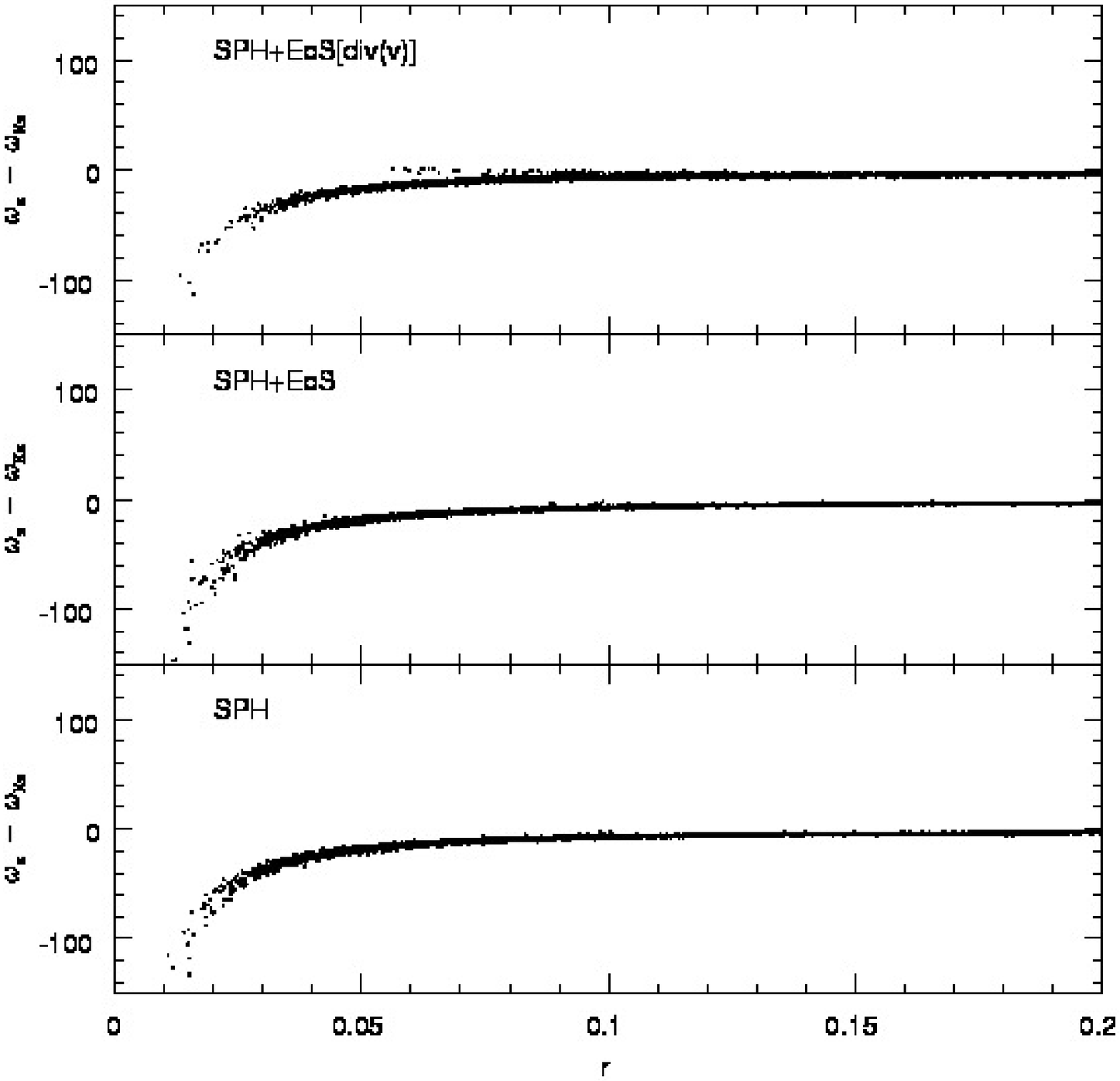}}
\caption{Radial distribution of deviation from the Keplerian kinematics, here represented as $\omega_{z} - \omega_{K}$, for the inner disc regions.}
\end{figure}

\begin{figure}
\resizebox{\hsize}{!}{\includegraphics[clip=true]{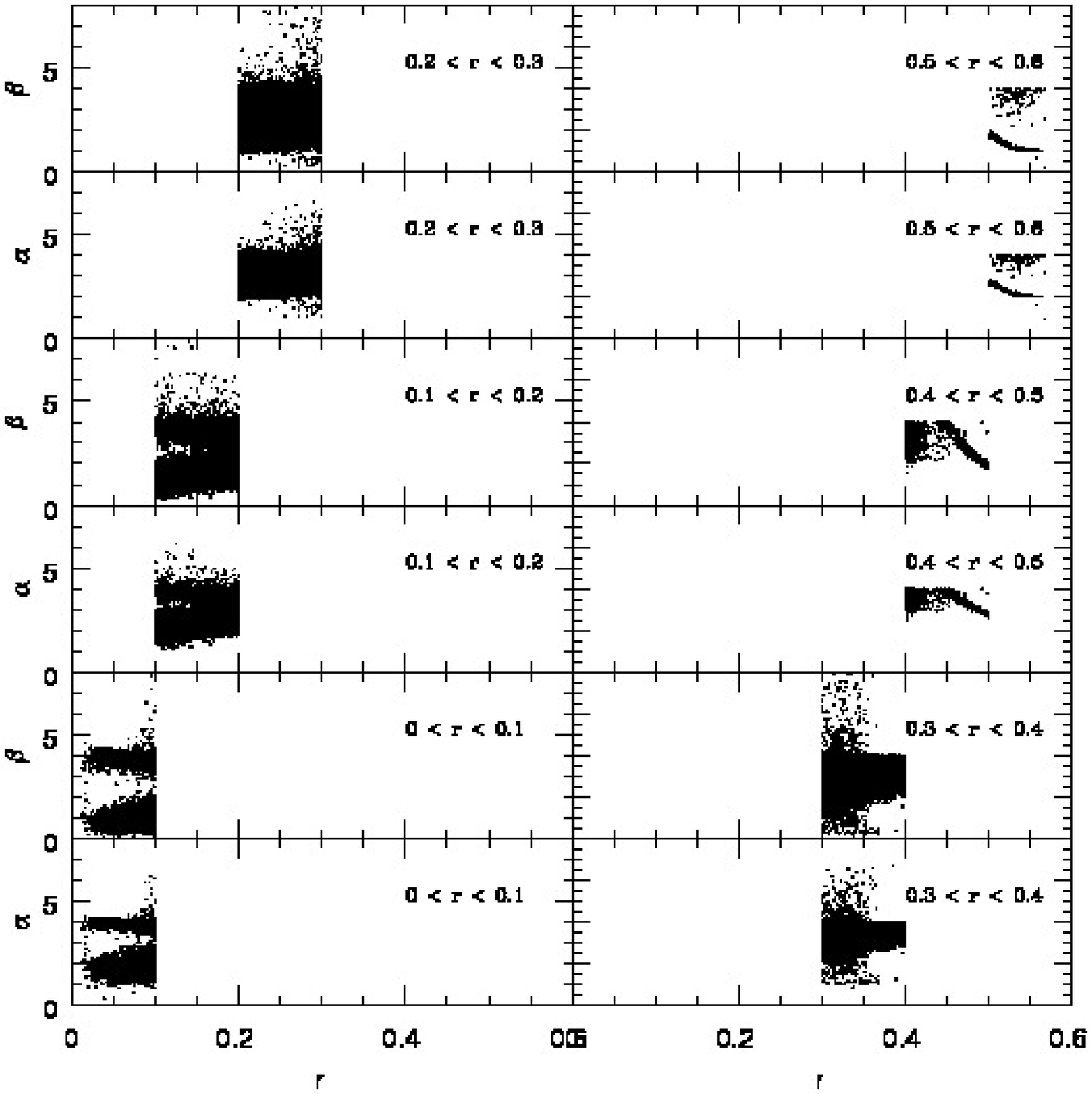}}
\caption{Radial distribution of the minimum and the maximum values of $\alpha$ and $\beta$ for each particle computed according to eqs. (42) and (43), respectively. Particles belonging to six $\Delta r = 0.1$ radial shells from $r = 0$ to $r = 0.6$ are represented.}

\resizebox{\hsize}{!}{\includegraphics[clip=true]{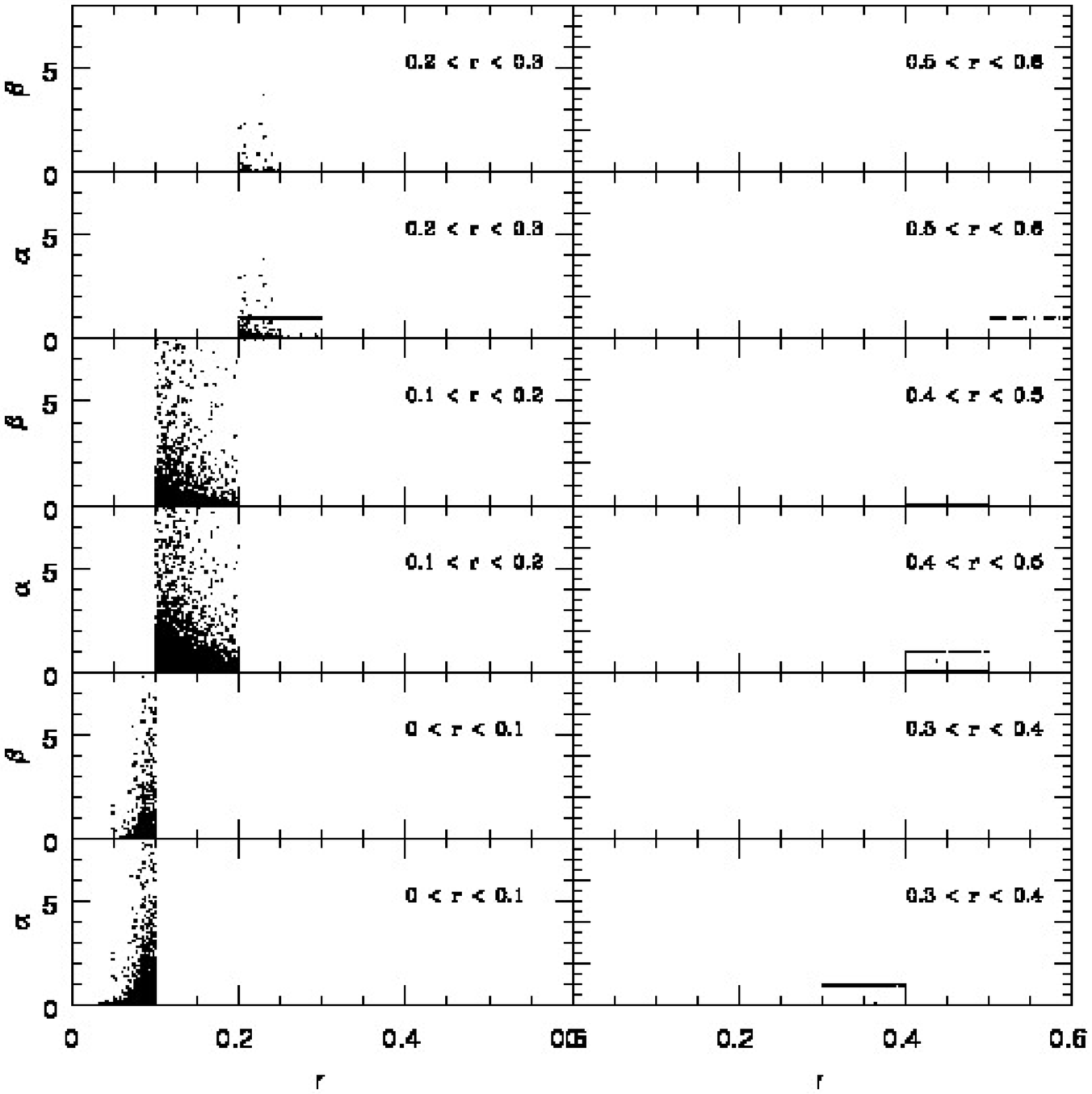}}
\caption{Radial distribution of the minimum and the maximum values of $\alpha$ and $\beta$ for each particle computed according to eqs. (62) and (63), respectively. Particles belonging to six $\Delta r = 0.1$ radial shells from $r = 0$ to $r = 0.6$ are represented.}
\end{figure}

  As an astrophysical application of EoS modification in fluid dynamics for a better identification of shock profiles, we pay attention to the development of spiral patterns in accretion discs in close binaries (CB), where collisions, shear flows, turbulence and vorticity are concurrent throughout the structure and dynamics of the flows. Some remarkable highlights are here written on this theme to show whether, how and why the accuracy in the Riemann problem can affect the quality of results in non viscous flows.

  Results on comparison of 3D SPH simulations of accretion discs in CB are here shown as a further check with the aim of showing which method, including an explicit dissipation inviscid term in the EoS, gives results where clear spiral patterns and shock fronts in the radial flow are evident. A rich scientific literature exists on spiral patterns appearance e.g. \citep{a34,a36,a37,a35,a38}. Each simulation is stopped when a steady state configuration is obtained: i.e. when the number of particles within the gravitational potential well of the primary compact star (e.g. a white dwarf or a neutron star) is statistically constant. The secondary star (a subgiant or a normal star) fills up entirely its Roche lobe, transferring its mass to the primary through the inner Lagrangian point L1. Disc's edges are free. \citet{a12,a13} clearly showed which geometric and kinematic conditions favour the appearance of spiral structures in SPH. Therefore, in this section we refer to those initial (at L1 point) and boundary conditions to highlight such structures. Particles move within the primary's Roche lobe towards the primary star at disc's inner edge (towards a central sphere whose radius equals $10^{-2}$), while they freely move outward from the disc's outer edge. Within the $10^{-2}$ radius they are considered in free fall, thus they are accreted. The smoothing resolution length of SPH particles is $h = 5 \cdot 10^{-3}$, being $1$ the non-dimensional separation $d_{12} = 10^{6} \ Km$ of the two stars. The injection of particles from L1 is supersonic: $v_{inj} = 130 \ Km \ s^{-1}$, whilst the temperature of gas coming from the secondary star is $T = 10^{4} \ K$ and $\gamma = 1.01$. The compact primary is a $1 M_{\odot}$ star, while the donor companion is a $0.5 M_{\odot}$ star.

  The adoption of supersonic mass transfer conditions from L1 is fully discussed in \citet{a43,a44}, where disc instabilities, responsible for disc active phases of CB are discussed in the light of local thermodynamics. Whenever a relevant discrepancy exists in the mass density across the inner Lagrangian point L1 between the two stellar Roche lobes, a supersonic mass transfer occurs as a consequence of the momentum flux conservation. The same result can also be obtained \citep{a41} by considering either the restricted problem of three bodies in terms of the Jacobi constant or the Bernoulli's theorem.

  The coming out of spiral patterns, as well as disc ellipticity are strongly evident in the disc model "SPH+EoS" of Fig. A1, where the EoS is reformulated according to a physical dissipation right for Riemann problem solution. The "SPH" disc model shows a glimmer of spiral structures, stimulated by supersonic injection conditions as evidenced in \citet{a12,a13}. Instead, the plot of Fig. A1, relative to the disc model "SPH+EoS[div(v)]", taking into account the full reformulation of the ideal gas equation in its non viscous physical sense, does not show such spiral characteristics. A central huge gas concentration, at a radial distance corresponding to the angular momentum conservation of particles injected from L1, is shown. Hence, the whole disc's structure and kinematics is dominated by a central toroidal Keplerian structure since a physical dissipation, ruled out by local compression of gas bubbles, is rarely activated. However, when a neighbour deviates from the strict Keplerian kinematics, it could be very strong. This demonstrates that the introduced physical dissipation within the EoS, suitable for the solution of the Riemann problem (eqs. 36, or 37+38, or 56), affects the particle kinematics, preventing a too strong circularization of gas orbits in strictly Keplerian orbits. Such an "Eulerian" physical dissipation, as well as any other physical viscous dissipation, coming from a Navier-Stokes viscous, is absolutely necessary if the appearance of spiral structures are the desired dominant characteristics of disc structure. The stronger the dissipation, the higher the contrast, within some limitations. The prevention of any spiral has been noted in \citet{a94}, modelling the same accretion discs, where an arbitrary attenuation of artificial viscosity, mainly on $\alpha$, has been applied according both to \citet{a15} and to \citet{a16}.
  
  The connection of disc ellipticity with spiral patterns is accurately discussed in \citet{a48}, where a third spiral pattern can also develop in some models. Typically the two main spirals are on opposite sides in the disc structure. The first one is directly connected to the incoming flow stream from L1, while the second one comes from the more elongated disc outer edge on the opposite side to the injected inflow. The newest third one comes from the more elongated disc outer edge close to the injected stream. Other previous high compressibility (low $\gamma$) SPH non viscous disc models in CB \citep{a1,a2,a3,a4,a12,a13,a18,a49} did not produce a such elliptical disc geometry. Flow perturbations, as well as tidal torques, are responsible for spiral appearance in disc structures. A very accurate discussion on tidal torque, its role and limits can be found in \citet{a56,a52,a53,a54,a55,a49}. In \citet{a49} we paid attention to the necessity to develop a Riemann-SPH code able to verify if weak shock fronts can develop within the disc bulk. After some time, we are now respecting that promise by considering a physical solution to the problem without any specific mathematical Riemann solver technique.
  
  Fig. A2 shows the radial distribution of the deviation of angular velocity component $\omega_{z}$ to the Keplerian angular velocity: $\omega_{z} - \omega_{K}$ in the inner disc regions. Of course, the "SPH+EoS[div(v)]" model shows the smaller deviation, both in its statistical sense (thickness of the distribution), and in the innermost radial regions, where the other two disc models progressively deviate, showing a lower angular momentum component and, therefore a higher accretion rate onto the compact primary star. Together with Fig. A1, this further result shows, without any doubt, that the EoS perfect gas reformulation (eq. 60) is really general because the "Eulerian" physical dissipation introduced is effective both in the shear flow fluid dynamics and in the Riemann problem flows at the same time.

  Fig. A3 shows the radial distribution of the minimum and the maximum values of $\alpha$ and $\beta$, respectively, according to eqs. (42, 43), if the EoS in its (36, or 37-38, or 56) forms is considered. Those extreme values are computed comparing $\alpha$ and $\beta$ for each neighbour for each particle. Both minimum and maximum $\alpha$ and $\beta$ values converge towards a single value $\approx 2$ and $\approx 1$, respectively at L1 due to the injector positioning. Spare higher $\alpha$ and $\beta values$ for $0.5 \leq r \leq 0.6$ refer to other disc regions at its outer edge. In the disc bulk, the two distinct domains, on average, enlarge being in contact in the boundary of $\approx 3.25$ and $\approx 2.75$, respectively, since high temperature radial gradients characterize the inner disc zones, so that both "much colder" and "much hotter" neighbour companions affect the physical dissipation. All other intermediate $\alpha$ and $\beta$ values, relative to each particle (not represented for the sake of simplicity), span within the two separated domains for each panel of Fig. A3. Such plots show that $\alpha$ and $\beta$ artificial viscosity counterpart of physical dissipation are larger, on the average, than the traditional $\alpha = 1$ and $\beta = 1 \div 2$, in so far as eqs. (36, or 37+38, or 56) are taken into account. However, because any dissipation is physically prevented when $\bmath{r}_{ij} \cdot \bmath{v}_{ij} \geq 0$, the $\alpha$ and $\beta$ counterparts are not always effective.
  
  Fig. A4 shows, as in the previous figure, the radial distribution of the minimum and the maximum values of $\alpha$ and $\beta$, according to eqs. (62, 63), if the EoS in its (60) form is attributed to each particle. Even in this case, this decision is taken to avoid too many points in the plots, representing each neighbour particle. The scale in Fig. A4 is the same as for Fig. A3 for the sake of simplicity. Both $\alpha \approx 0$ and $\beta \approx 0$ in the disc's outer regions, free of any gas compression, where local kinematics is mainly Keplerian at the disc's outer edge. This means that $\nabla \cdot \bmath{v}$ is locally negligible. On the other external components, as the injected particle stream, no local compression occurs. For radial distance $r \leq 0.2$, where the majority of disc's particles relies in the inner toroidal ring, both $\alpha$ and $\beta$ increase since the kinematics deviates from a perfect Keplerian behaviour and gas compression becomes effective there. This means that a non viscous hydrodynamics of a perfect gas, without a shear physical dissipation, also effective without any gas compression, cannot produce any graduality in the radial transport. This implies that in an ideal gas, either the turbulent physical dissipation in a Navier-Stokes approach, or an inviscid Eulerian fluid dynamics, where the Riemann problem is correctly solved, must be considered to produce both a graduality in the radial transport, and spiral shaped structures and a statistically correct Keplerian tangential kinematics in a disc.

\label{lastpage}

\end{document}